\documentclass[aps,prb,twocolumn,amsfonts,amsmath,amssymb,superscriptaddress,showpacs]{revtex4-1}

\usepackage{graphicx}
\usepackage[utf8]{inputenc}
\usepackage{listings}
\usepackage{bm}
\usepackage{setspace}
\usepackage{hyperref}
\hypersetup{colorlinks=true,citecolor=blue,urlcolor=blue,linkcolor=blue,breaklinks=true}
\usepackage{natbib}
\usepackage[normalem]{ulem}
\usepackage{epstopdf}
\usepackage{color}
\usepackage{dsfont}
\usepackage[bbgreekl]{mathbbol}
\usepackage{amsfonts}
\usepackage{amsmath, amssymb}
\usepackage{amsthm}
\usepackage{soul}

\def\d{\mathrm{d}}

\usepackage[utf8]{inputenc}
\usepackage[english]{babel}
\usepackage{amsmath}
\usepackage{amssymb}
\usepackage{mathtools}
\usepackage{algorithm}
\usepackage{algorithmic}
\usepackage[binary-units=true]{siunitx}
\usepackage{enumerate}
\usepackage{soul}
\usepackage{cleveref}

\usepackage{fullpage}
\usepackage{booktabs}
\usepackage{array}

\usepackage{tikz}
\usepackage{xcolor}
\usetikzlibrary{patterns}
\usetikzlibrary{arrows,decorations.pathreplacing}

\usepackage{url}

\renewcommand{\d}[1]{\textrm{ d}#1}

\newcommand{\iu}{\mathrm{i}\mkern1mu}
\newcommand{\bra}[1]{\langle#1|}
\newcommand{\ket}[1]{|#1\rangle}

\newcommand{\operator}[1]{\hat{#1}}
\newcommand{\creation}[1]{\operator{c}^\dagger_{#1}}
\newcommand{\annihilation}[1]{\operator{c}^{\phantom\dagger}_{#1}}
\newcommand{\occupation}[1]{\operator{n}_{#1}}
\newcommand{\doubleoccupation}[1]{\operator{d}_{#1}}
\newcommand{\spinup}{\uparrow}
\newcommand{\spindown}{\downarrow}
\newcommand{\nsites}{{N_s}}

\newcommand*\colvec[3][]{
    \begin{pmatrix}\ifx\relax#1\relax\else#1\\\fi#2\\#3\end{pmatrix}
}

\makeatother
\begin{document}
\noindent

\title{Enhancement of impact ionization in Hubbard clusters by disorder and next-nearest-neighbor hopping}

\author{Anna Kauch}
\affiliation{Institute for Solid State Physics, Vienna University of Technology,  Wiedner Hauptstr. 8-10, 1040 Wien, Austria}

\author{Paul Worm}
\affiliation{Institute for Solid State Physics, Vienna University of Technology,  Wiedner Hauptstr. 8-10, 1040 Wien, Austria}

\author{Paul Prauhart}
\affiliation{Institute for Solid State Physics, Vienna University of Technology,  Wiedner Hauptstr. 8-10, 1040 Wien, Austria}

\author{Michael Innerberger}
\affiliation{ Institute for Analysis and Scientific Computing, Vienna University of Technology, Wiedner Hauptstr. 8-10, 1040 Wien, Austria }

\affiliation{Institute for Solid State Physics, Vienna University of Technology,  Wiedner Hauptstr. 8-10, 1040 Wien, Austria}

\author{Clemens Watzenb\"ock}
\affiliation{Institute for Solid State Physics, Vienna University of Technology,  Wiedner Hauptstr. 8-10, 1040 Wien, Austria}

\author{Karsten Held}
\affiliation{Institute for Solid State Physics, Vienna University of Technology,  Wiedner Hauptstr. 8-10, 1040 Wien, Austria}

\date{\today}

\begin{abstract}
We perform time-resolved exact diagonalization of the Hubbard model with time dependent hoppings on small clusters of up to $12$ sites.  Here, the time dependence originates from a classic electromagnetic pulse, which mimics the impact of a photon. We investigate the behavior of the double occupation and spectral function after the pulse for different cluster geometries and on-site potentials. We find impact ionization in all studied geometries except for one-dimensional chains.
Adding next-nearest neighbor hopping to the model leads to a significant enhancement of impact ionization, as does disorder and geometric frustration of a triangular lattice. 

\end{abstract}


\maketitle


\section{Introduction}

A more efficient solar energy conversion is urgently sought-after. Conventional semiconductor solar cell are however limited by the Shockley-Queisser limit \cite{shockley_queisser} of 34\% efficiency. A different class of materials has been shown to possess the  potential to overcome this barrier~\cite{manousakis2010,assmann2013,werner2014,sorantin2018}: transition metal oxides. These materials are at integer filling mostly Mott insulators with strong electronic correlations that result in a spectral gap. Such a gap is necessary for the active, photon-absorbing region of a solar cell. In Mott insulators a phenomenon called impact ionization may take place~\cite{manousakis2010,werner2014,sorantin2018}, which makes it possible to produce more than one electron-hole pair per photon. The time scale of impact ionization can be a few femtoseconds and hence several orders of magnitude faster than the electron-phonon processes in these materials (typically of the time scale of picoseconds). This is in strong contrast to semiconductor solar cells where impact ionization is of the order of several  picoseconds \cite{Shockely61,Keldysh65,shockley_queisser} so that any excess kinetic energy of the photon-generated electron-hole pair is absorbed as thermal lattice vibrations, instead of producing electrical energy.  This leads to the  Shockley-Queisser limit in the first place. Besides the prospects of impact ionization, transition metal oxides can be produced as hetrostructures with a potential gradient at a polar interface. This gradient or the corresponding field  enables an electron-hole separation and allows one to harvest  the excess charge in form of a current~\cite{assmann2013,sorantin2018,Kropf2020,Petocchi2019}. Experimentally such transition metal oxide heterostructures have been demonstrated to act as solar cells, on the basis of the Mott insulators   LaVO$_3$\cite{Wang2015,Zhang2017} and LaFeO$_3$ \cite{Nakamura2016}.

In this paper we focus only on the phenomenon of impact ionization, that is creation of an additional electron-hole pair, after photo-excitation of the first one, due to electron-electron interaction. It has been studied theoretically in the Hubbard model on a lattice  with dynamical mean-field theory~\cite{werner2014,sorantin2018}, using Fermi's golden rule \cite{manousakis2010,Manousakis2019} and the Boltzmann equation~\cite{wais2018}. Experimentally, evidence for impact ionization was demonstrated in VO$_2$, \cite{Holleman2016} and in  quantum dots \cite{Franceschetti06,Wang13,Wang17}.  

Here, we provide a complementary theoretical approach: instead of studying a non-equilibrium extended system with an approximate method, we use exact diagonalization of the Hubbard model on a small cluster of sites~\cite{innerberger2020,Maislinger2020}. The interaction with light is modeled by adding a time dependent classical light pulse to the Hamiltonian via Peierls' substitution. We confirm that impact ionization is present in clusters as small as $8$ sites and analyze its dependence on model parameters and geometry. Surprisingly, we find that disorder does not damage the effect, but leads to an enhancement of impact ionization. While in part this effect may be specific to the small system sizes we are able to study  (up to $12$ sites), our study certainly provides an incentive to study the effects of disorder  in extended systems with impact ionization. 

As far as geometry and connectivity is concerned, we find that the number of neighbors is important -- the more neighbors available for the electrons to hop to, the stronger the impact ionization. We do not find impact ionization in chain geometries with only nearest neighbor hopping. Apart from a small number of neighbors, the 1D systems are hosting  strong antiferromagnetic spin fluctuations~\cite{Essler05}, which may be disfavorable for impact ionization. Indeed, for a $10$-site fragment of a triangular lattice, which is magnetically frustrated, we find quite strong impact ionization.     

As the basic measure for impact ionization we take the increase of double occupation as a function of time at times after the light pulse, when also the total energy does not change any more. Then, the increase of double occupation can only be caused by impact ionization, since photo-excitation is no longer possible. We also calculate the non-equilibrium time dependent spectral function and analyze the time evolution of spectral weight in the Hubbard bands. The impact ionization makes itself visible in the spectral weight shift both inside the upper Hubbard band as well as from the upper Hubbard band to the lower Hubbard band occurring at times after the pulse. We also see photo-induced  gap filling  (photo-melting of the Mott-insulator) that is stronger in cases where impact ionization is also stronger. The phenomenon of light-induced gap filling has also been reported in~\cite{wais2018,okamoto2019,innerberger2020}. In equilibrium a related filling of the Mott-Hubbard gap occurs at elevated temperature. \cite{Mo2004}

The paper is organized as follows: In Sec.~\ref{Sec:Model} we introduce the model, notation and units used throughout the paper. We also describe here the different geometries we study. In Sec.~\ref{Sec:Results} we present our results for the double occupation and spectral function during and after the electric field pulse. Different geometries of the Hubbard clusters are studied:
chains and boxes with nearest neighbor hopping only in {Sec.~}\ref{Sec:NN},
additional next-nearest neighbor hopping in {Sec.~}\ref{Sec:NNN}, a triangular geometry in {Sec.~}\ref{Sec:Tri}, and the effect of disorder in {Sec.~}\ref{Sec:disorder}. Finally, we identify common trends for the different geometries in {Sec.~}\ref{Sec:Disc} and summarize our main findings in {Sec.~}\ref{Sec:Summary}. Additional plots are also presented in the Appendix.

\section{Model and method}
\label{Sec:Model}

The paradigm model for studying strongly interacting electrons is the Hubbard model~\cite{hubbard1963}, given by the following Hamiltonian
\begin{equation}
\operator{H} = \sum_{i,j,\sigma} v_{ij} \creation{j\sigma}\annihilation{i\sigma} + U\sum_{i}n_{i\uparrow}n_{i\downarrow}
\label{eq:Hubbard}.
\end{equation}
Here, $\creation{i\sigma}$ ($\annihilation{i\sigma}$) creates (annihilates) an electron on site $i$ with spin $\sigma$, $n_{i\sigma}=\creation{i\sigma}\annihilation{i\sigma}$ is the occupation number operator,  $U>0$ is the  local Coulomb repulsion, and $v_{ij}$ for $i\neq j$ describes the hopping amplitude from site $i$ to $j$ and for $i=j$ an additional, site dependent on-site potential. In the following we will restrict the hopping to either nearest-neighbor (NN) or NN and next-nearest-neighbor (NNN) sites. We also choose the system to be half-filled with the number of electrons with either spin given by $\nsites/2$  (with $\nsites$ being the number of sites).

\subsection{Peierls' substitution}

\label{sec:peierls}
The light is modeled as a classical electric field pulse \cite{werner2014}
\begin{equation}\label{eq:efield}
	\vec{E}(t)=\vec{E}_0\sin(\omega_p(t-t_p))e^{-\frac{(t-t_p)^2}{2 \sigma^2}}
\end{equation}
of width $\sigma$, peaked around the time $t_p$, and with frequency $\omega_p$. We set the units of frequency equal to the units of energy ($\hbar \equiv 1$) and the unit of time is the inverse of the unit of energy. For typical values of the NN hopping in transition metal oxides $\sim 0.3$eV, the unit of time is around $2$fs.

The electromagnetic (EM) field is included in the Hubbard Hamiltonian using Peierls' substitution~\cite{Peierls}, which adds a time dependence to the hoppings:
\begin{equation} \label{eq:peierls}
	v_{ij} \rightarrow v_{ij}(t)=v_{ij}\exp\left({-\iu e \int_{\vec{R}_i}^{\vec{R}_j}\vec{A}(\vec{r}',t)d\vec{r}'}\right).
\end{equation}
We use a gauge where the scalar potential vanishes and $\vec{E}=-\partial_t\vec{A}(t)$. The wavelength of light is assumed to be much longer than the system size, which renders $\vec{A}$ only time dependent. The value of the integral in~\eqref{eq:peierls} for different pairs of sites $i$ and $j$ depends on $\vec{A}\cdot (\vec{R}_j-\vec{R}_i)$. The time dependence has the same form for all hoppings. The vector potential $\vec{A}(t)$ is obtained by integrating the  $\vec{E}$-field in~\eqref{eq:efield}, and can be further approximated by only integrating the sine function in~\eqref{eq:efield} if the light pulse contains many $\omega_p$  oscillations (i.e., $1/\omega_p \ll \sigma$). Then we arrive at the following time dependence of the hoppings:
\begin{equation} \label{eq:vt}
v_{ij}(t)=v_{ij}\exp\left(\iu g_{ij}a\left[\cos(\omega_p(t-t_p))-b\right]e^{-\frac{(t-t_p)^2}{2\sigma^2}}\right),
\end{equation}
where $g_{ij}$ is a dimensionless parameter that depends only on geometry and is given by the relative angle between $\vec{A}$ and $\vec{R}_j-\vec{R}_i$. The dimensionless parameter $a$ describes the strength of the EM field, whereas $b$ can be used to set the initial phase factor of the hoppings to $1$. Note, that the Peierls' substitution introduces only a phase factor to the hoppings and does not change their absolute value. For all results presented in this paper, the NN hoppings will be set to have equal absolute value and this hopping amplitude  is used as the unit of energy, i.e.\ $|v_{ij}| = 1$.

\subsection{Geometry}
\label{sec:geometry}
\begin{figure}
	\centering
	\includegraphics[width=0.9\linewidth]{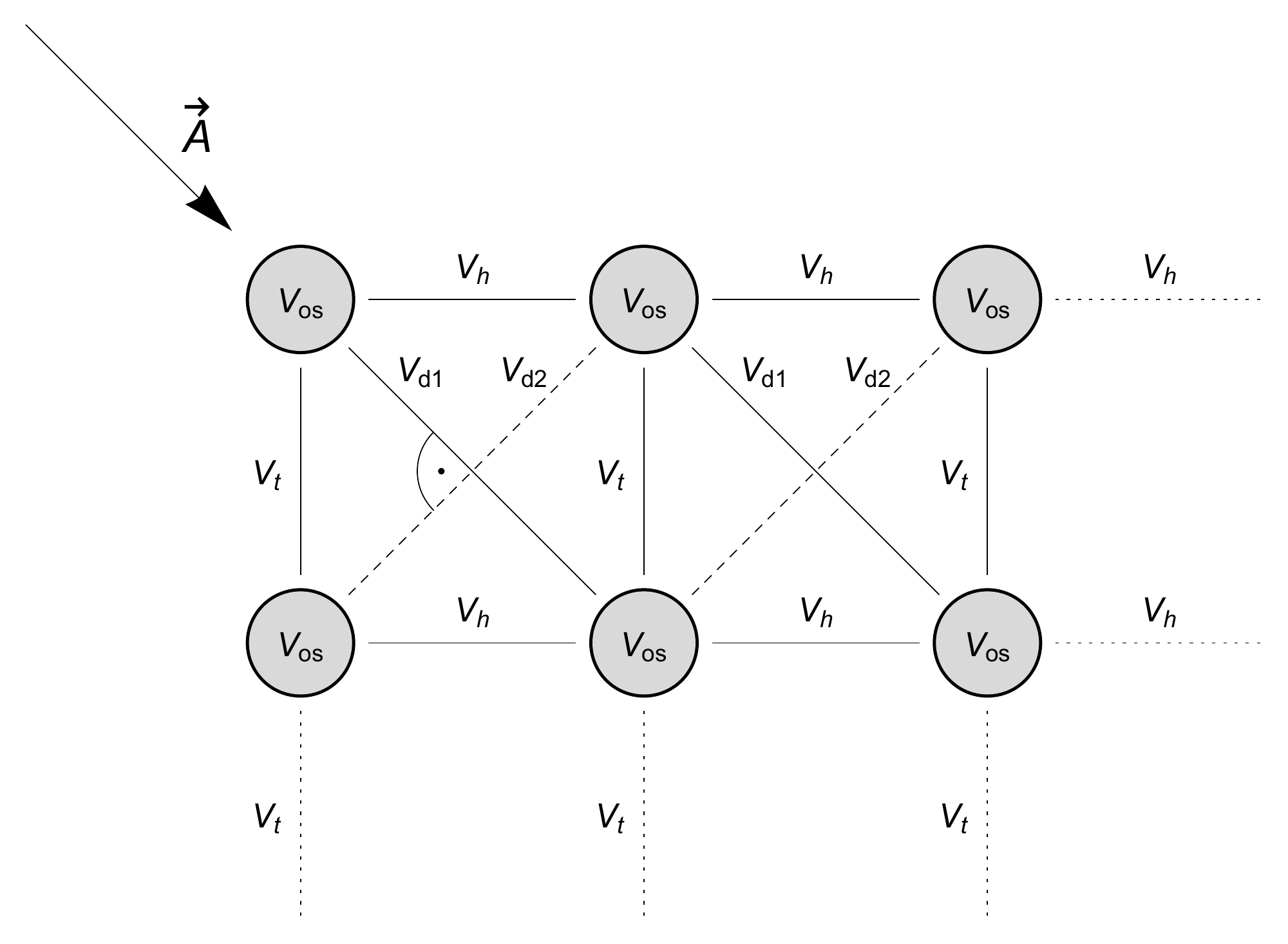}
	\caption{Example of a  $2\times 3$ box geometry with  on-site potential equal on all sites $v_{ii}=v_{os}$, NN hopping $v_h$ in horizontal direction and $v_t$ in the vertical direction as well as two different NNN (diagonal) hoppings $v_{d1}$ and  $v_{d2}$. In our simulations, the time-independent prefactors are equal for NN hoppings $v_t=v_h\equiv 1$ and  for NNN hoppings $v_{d1}=v_{d2}\equiv v_d$ .
If the vector potential $\vec{A}$ is chosen along one of the diagonal directions (as shown in the Figure and employed in our calculation), the geometric factor $g_{ij}$ determining the time dependence in ~\eqref{eq:vt} is the same for $v_h$ and $v_t$, it is twice as big for $v_{d1}$ and zero for $v_{d2}$.}
	\label{fig:geometry}
\end{figure}

The information about the geometry of the system is entirely given by the elements of the hopping matrix $v_{ij}$. We use open boundary conditions and chain or box geometries. The distance between sites is taken to be equal in the horizontal and vertical direction. The direction of the vector potential $\vec{A}$ is chosen to create a $45^{\rm o}$ angle with the horizontal direction, as sketched in Fig.~\ref{fig:geometry} for a $2\times 3$ box. This way the geometric factor $g_{ij}$ needed in Eq.~\eqref{eq:vt} is the same for vertical and horizontal NN hoppings, twice as big for NNN hopping parallel to  $\vec{A}$ and zero for NNN hopping perpendicular to $\vec{A}$.

\subsection{Disorder}
\label{Sec:Disorder}
To study the effects of site disorder we use a set of uniformly distributed random numbers (box disorder):
\begin{equation}\label{eq:disorder}
\epsilon_i \in (-\Delta/2, \Delta/{2}), \quad i=1.\ldots \nsites
\end{equation}
which shift the on-site potentials from their particle-hole symmetric value of $-U/2$, i.e.,
\begin{equation}\label{eq:onsite}
v_{ii}=-\frac{U}{2} + \epsilon_i.
\end{equation}
In all plots presented in Sec.~\ref{Sec:Results} the strength of disorder is given as a percentage of the Coulomb interaction $U$, i.e. as the ratio $\Delta/U\cdot 100\%$. The results are averaged (with arithmetic averaging) over the disorder realizations.

\subsection{Time evolution}
\label{Sec:timeevolution}
In order to calculate the time evolution of the system driven out of equilibrium by a time-dependent light pulse, we solve the time-dependent Schr\"odinger equation
\begin{equation}
\label{eq:schroedinger}
\iu \partial_t \ket{\psi(t)} = \operator{H}(t) \ket{\psi(t)},
\qquad \ket{\psi(0)}=\ket{\psi_0}
\end{equation}
using a time-stepping algorithm described in detail in~Ref.~\onlinecite{innerberger2020}, for a finite system of $\nsites$ sites. The initial state $\ket{\psi_0}$ of the time evolution is always taken to be the ground state. The time is discretized and for each time step $\delta t$ the midpoint rule
\begin{equation}
\label{eq:midpoint}
	\ket{\psi(t+\delta t)} \approx \exp\left( -\iu \delta t H(t+\delta t/2)  \right) \ket{\psi(t)}
\end{equation}
is applied. For higher order Magnus integrators applied to similar problems see Ref.~\onlinecite{magnus_paper}. The resulting matrix exponentiation is performed using Krylov subspace method~\cite{expokit}. We used the value of $\delta t =0.005$ for all computations in this paper.

\subsection{Impact ionization}
\label{Sec:impact}
The phenomenon of impact ionization can be understood as follows: with a fixed quantum of photon energy, the excited electrons and holes (or doublons and holons) have excess  kinetic energy. If the photon energy is larger than twice the Mott gap, it is possible to convert the excess kinetic energy  through a process coined impact ionization\cite{Shockely61,Keldysh65,ATOMimpact} into potential energy of one (or more) additional electron-hole pairs. The easiest way to observe if such processes take place is to look at the time dependent double occupation (the potential energy in the Hubbard model is just given by double occupation multiplied by $U$). In the following we hence compute the time-dependent site-averaged double occupation:
\begin{equation}\label{eq:docc}
\langle\doubleoccupation{}(t)\rangle = \tfrac{1}{\nsites} \sum_{i=1}^{\nsites}\bra{\psi(t)}\doubleoccupation{i}\ket{\psi(t)},
\end{equation}
with $\doubleoccupation{i}=\occupation{i\spinup}\occupation{i\spindown}$.

Another source of physical information about the system is the part of the spectral function that describes the occupied states $A^<(\omega,t)$. In equilibrium it is time independent and given by the spectral function $A(\omega)$ multiplied by the Fermi-Dirac distribution function $f_{FD}(\omega)$. In nonequilibrium we obtain  $A^<(\omega,t)$ by a forward Fourier transform~\cite{werner2016} of the lesser Green's function
\begin{equation}
    A_{ij\sigma}^<(\omega,t) =  \frac{1}{\pi} \text{Im} \int_0^{\infty} e^{\iu \omega t_{\text{rel}}} G_{ij\sigma}^<(t,t+t_{\text{rel}}) \d{t_{\text{rel}}}.
\label{eq:A_noneq}
\end{equation}
Having obtained the time evolution of the system  $\ket{\psi(t)}$ from Eq.~\eqref{eq:schroedinger}, we can calculate $G^<$ directly from 
\begin{equation}
    \begin{split}
        &G_{ij\sigma}^{<}(t,t') = \iu  \bra{\psi(t')}\creation{j\sigma} \mathcal{T} e^{-\iu\int_{t}^{t'} H(\tau) \d{\tau}} \;\annihilation{i\sigma}  \ket{\psi(t)} ,
    \end{split}
\label{eq:Glg}
\end{equation}
where $\mathcal{T}$ is the time ordering operator and the time evolution between times $t$ and $t'$ is calculated with the same time-stepping algorithm as $\ket{\psi(t)}$. Since the spectrum is discrete in $\omega$ for finite systems, we multiply $G^<$ in~\eqref{eq:A_noneq} with a broadening function $e^{-\epsilon t_{\text{rel}}}$ in our numerical implementation, which translates to a Lorentzian broadening in frequencies. The maximal value of $t_{\text{rel}}$ for the numerical evaluation of the integral in \eqref{eq:A_noneq} was $t^{\text{max}}_{\text{rel}}\approx80$.

In the following we will be interested in the site-averaged lesser part of the local spectral function:
\begin{equation}\label{eq:Alocal}
 A^<(\omega,t)= \tfrac{1}{\nsites} \sum_{i=1}^{\nsites}A_{ii\sigma}^<(\omega,t),
\end{equation}
which is identical for both spin directions here.

\section{Results}
\label{Sec:Results}

\begin{figure}
\centering
		 \includegraphics[width=0.6\linewidth]{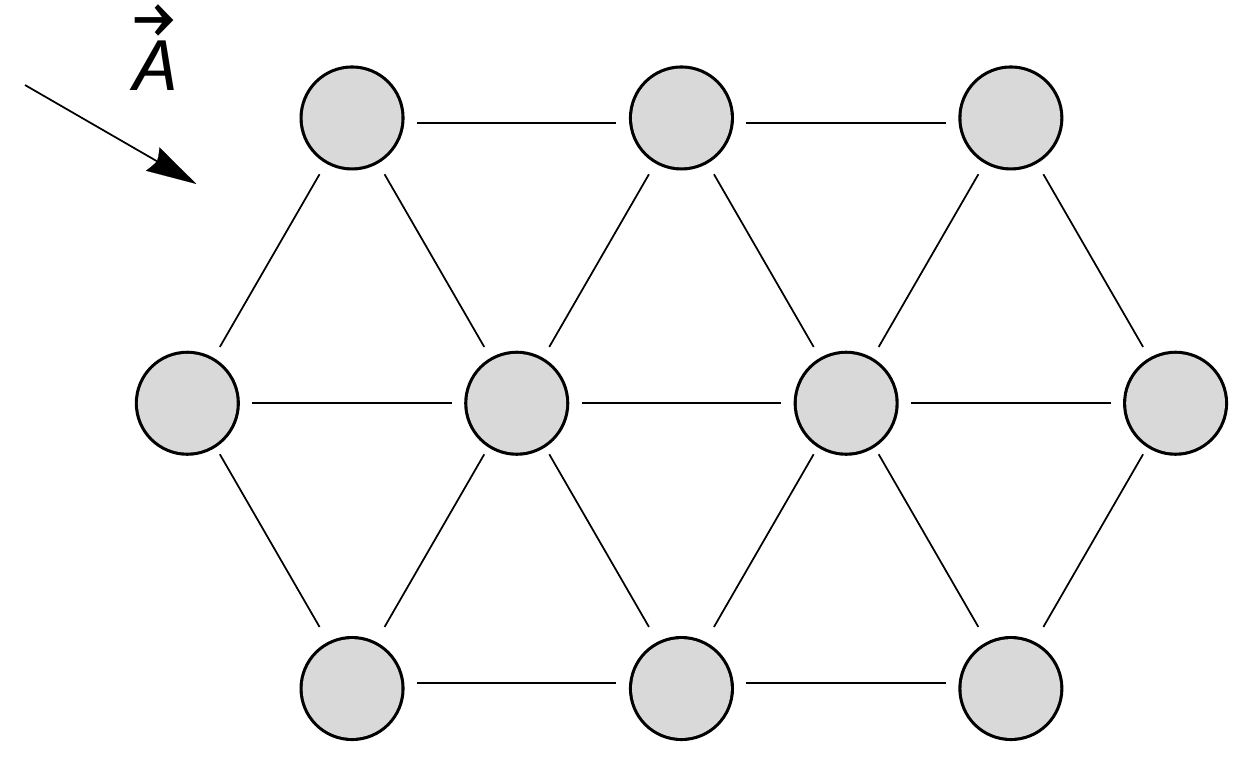}
		 \caption{The geometry of the $\nsites=10$ site cluster (denoted as tri-$10$). The sites are taken to be equidistant with equal hopping amplitude.}
		 \label{fig:tri10}
\end{figure}

In the following we present results obtained for several cluster sizes $\nsites$ and geometries, namely:
\begin{itemize}
\item[(i)] {Chains of length $\nsites=8$ and  $\nsites=12$ sites with only NN hopping, with or without disorder.}
\item[(ii)] {Boxes (two-dimensional rectangular clusters) of sizes $2\times 4$, $2\times 6$ and $4\times3$ with only NN or NN and NNN hopping, as depicted in Fig.~\ref{fig:geometry} (the strength  of the NNN hopping is denoted by $v_d=|v_{d1}|=|v_{d2}|$ in the following), with or without site disorder.}
 \item[(iii)] {A cluster $\nsites=10$ sites as depicted in Fig.~\ref{fig:tri10}, with NN  hopping, which we denote as tri-$10$ cluster in the following, as this cluster can be envisaged as  part of a triangular lattice.}
\end{itemize}

In all cases we use open boundary conditions and half-filling with the on-site potentials either equal to $-U/2$ (no disorder) or modified according  to Eqs.~\eqref{eq:disorder}-\eqref{eq:onsite} if we consider disorder. The parameters of the model and the pulse (interaction $U$, pulse frequency $\omega_p$, and intensity $a$) are chosen so that the effects discussed are most pronounced (for the $2\times4$ cluster we present a parameter scan in Appendix~\ref{app:scan} and for the $12$-site clusters we used the parameters from~\onlinecite{note_parameters}). The center of the pulse is set to $t_p=5$, the width to $\sigma=2$ and the time step is $\delta \tau =0.005$ (for more computational details see Ref.~\onlinecite{innerberger2020}, where the same time-stepping algorithm was applied). We always start the time evolution from the ground state, which is an insulator for all geometries and disorders.

\begin{figure}[t]
	\centering
 \includegraphics[width=\linewidth]{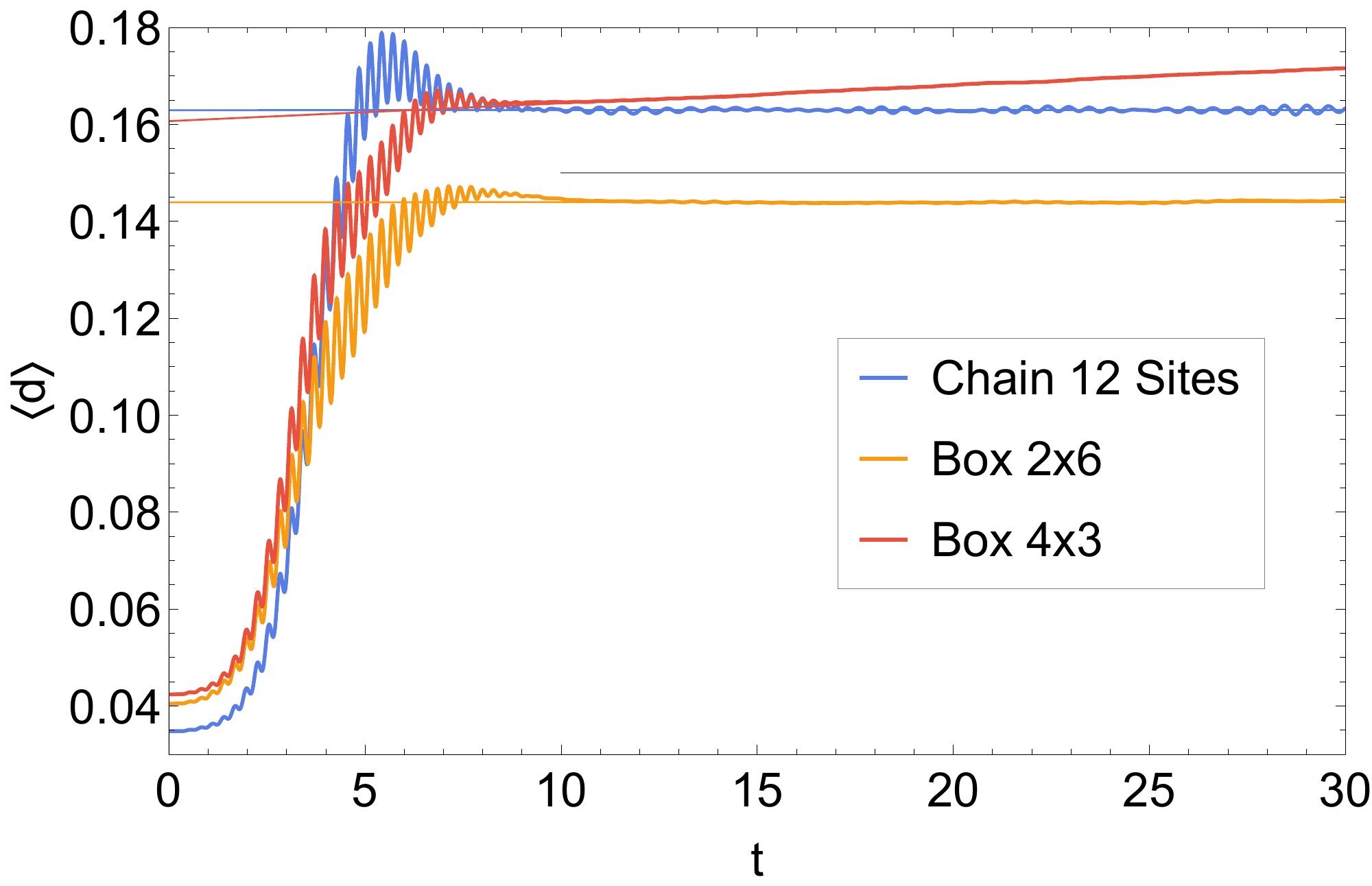}
		 \caption{Average double occupation as a function of time for a $12$-site chain, a $2\times6$ box and a $4\times3$ box at $U=8$,  $\omega_p=11$ and $a=0.8$. The blue, red and orange straight lines are linear fits to the data in the range $t \in [10,30]$. A horizontal gray line is  added to better visualize the small slope of the yellow curve. }
		 \label{fig:dspec1}
\end{figure}

\begin{figure}[t]
	\centering
 \includegraphics[width=\linewidth]{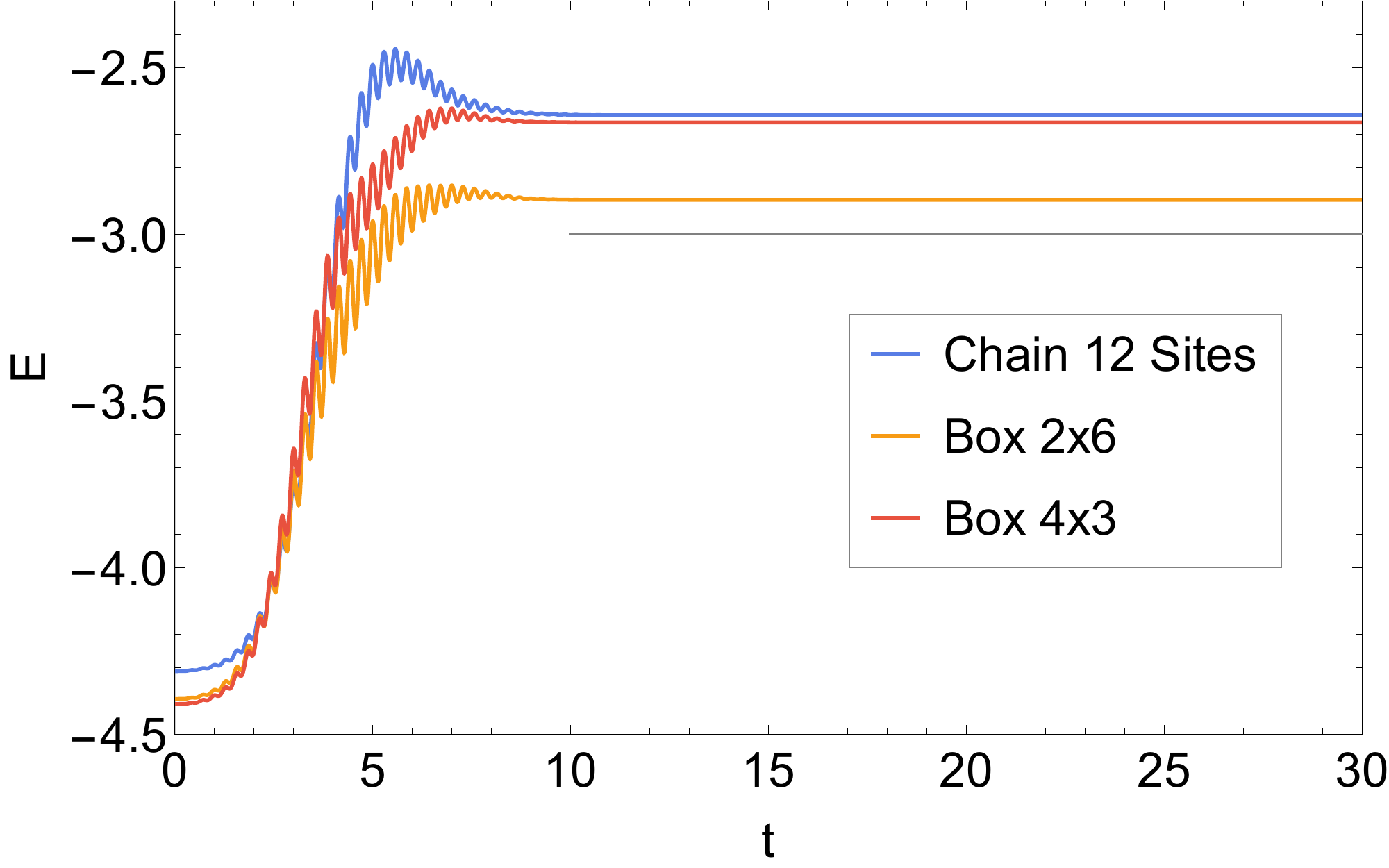}
		 \caption{Energy per site as a function of time for a $12$-site chain, a $2\times6$ box and a $4\times3$ box. The same parameters as in~Fig.~\ref{fig:dspec1}. }
		 \label{fig:espec1}
\end{figure}

\subsection{Systems with only NN hopping}
\label{sec:nn_only}
\label{Sec:NN}

We begin with presenting in Fig.~\ref{fig:dspec1} the double occupation as a function of time for three $12$-site systems: a $12$-site chain, $2\times6$ box, and $4\times3$ box, with only NN hopping and no disorder. The double occupation rises significantly during the pulse  at $t_p\pm \sigma=5\pm 2$ for all three systems: Energy is pumped into the system and electron-hole (or doublon-holon) pairs are created. For the  $4\times3$ box we see an additional rise of $\langle\hat{d}\rangle$ for later times, after the pulse is switched off, and the total energy does not change any more (cf. the time dependence of total energy per site shown in~Fig.~\ref{fig:espec1}). This further rise of potential energy (which is proportional to double occupation) is due to impact ionization. Electrons initially excited across the gap to the upper Hubbard band  have excess high kinetic energy. This excess kinetic energy is reduced by creating further electron-hole pairs~\footnote{At later times in realistic systems the electron-phonon interaction would be important for relaxation. Since we have only electronic effects in our model, we do not consider longer times and restrict to $t\le30$. In Ref.~\onlinecite{Maislinger2020} also longer times (up to 300) are shown.}. This phenomenon has already been reported for extended systems~\cite{werner2014,wais2018} and was also observed for the $4\times3$ box by Maislinger and Evertz~\cite{Maislinger2020}. We do not find any impact ionization in chain geometries (we have investigated chains up to $14$ sites). For the $2\times6$ box we see only a tiny rise of the double occupation. Since the double occupation fluctuates over time, we added linear fits to the data in Fig.~\ref{fig:dspec1}. The fitting was done only for times  after the pulse, $t \in [10,30]$, when the total energy of the system did not change any more. For the $12$-site chain the fit gives a horizontal line (the linear coefficient $k<10^{-6}$).

\begin{figure}
		\centering
		 \includegraphics[width=\linewidth]{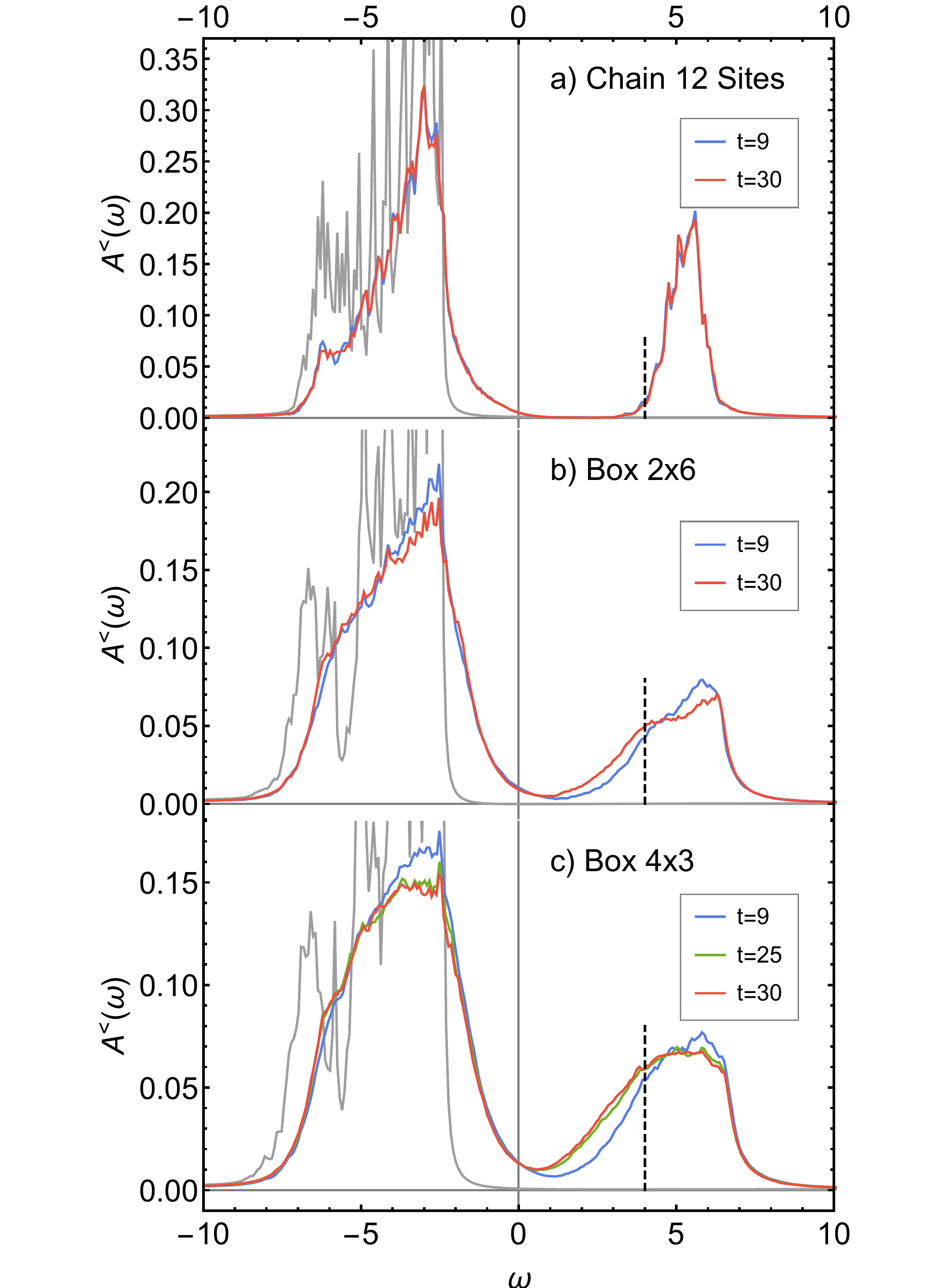}
		 \caption{$A^<(\omega,t)$  at three different times $t$  for the a)  $12$-site chain,  b)  $2\times 6$ box,  and c) $4\times3$ box; with the same parameters as in Fig.~\ref{fig:dspec1}. The gray curve is $A^<(\omega)$ in the ground state.  The dashed line indicates the separation into lower and upper part of the UHB used in Fig. \ref{fig:uhb1}. The results are broadened with $\epsilon$=0.04.	 \label{fig:spec1}}
\end{figure}

In Fig.~\ref{fig:spec1} we show the corresponding (site averaged) $A^<(\omega,t)$ for the three $12$-site systems for several different times after the pulse: $t=10$, $t=25$, $t=30$. Physically,  $A^<(\omega,t)$ corresponds to a momentum integrated photoemission spectroscopy, under certain approximations.
The most visible difference between the plots is that for the chain geometry the spectral function remains practically unchanged, whereas for the other two systems there are significant spectral weight shifts inside the Hubbard bands and in the $4\times3$ system also between the bands. A similar redistribution of spectral weight for an extended Bethe lattice has been found in Ref.~\onlinecite{werner2014}. 
To quantify the weight shifts inside the upper Hubbard band (UHB), we separate it into the lower part of UHB ($\omega \in [0,4]$) and upper part of UHB ($\omega \in (4,8]$). The division point of $\omega=4$, indicated in Fig.~\ref{fig:spec1} by a vertical dashed line, is chosen approximately in the middle of the UHB. The conclusions we make do not dependent on its specific choice, as long as it is close to the middle of the UHB. 

\begin{figure}
		\centering
		 \includegraphics[width=\linewidth]{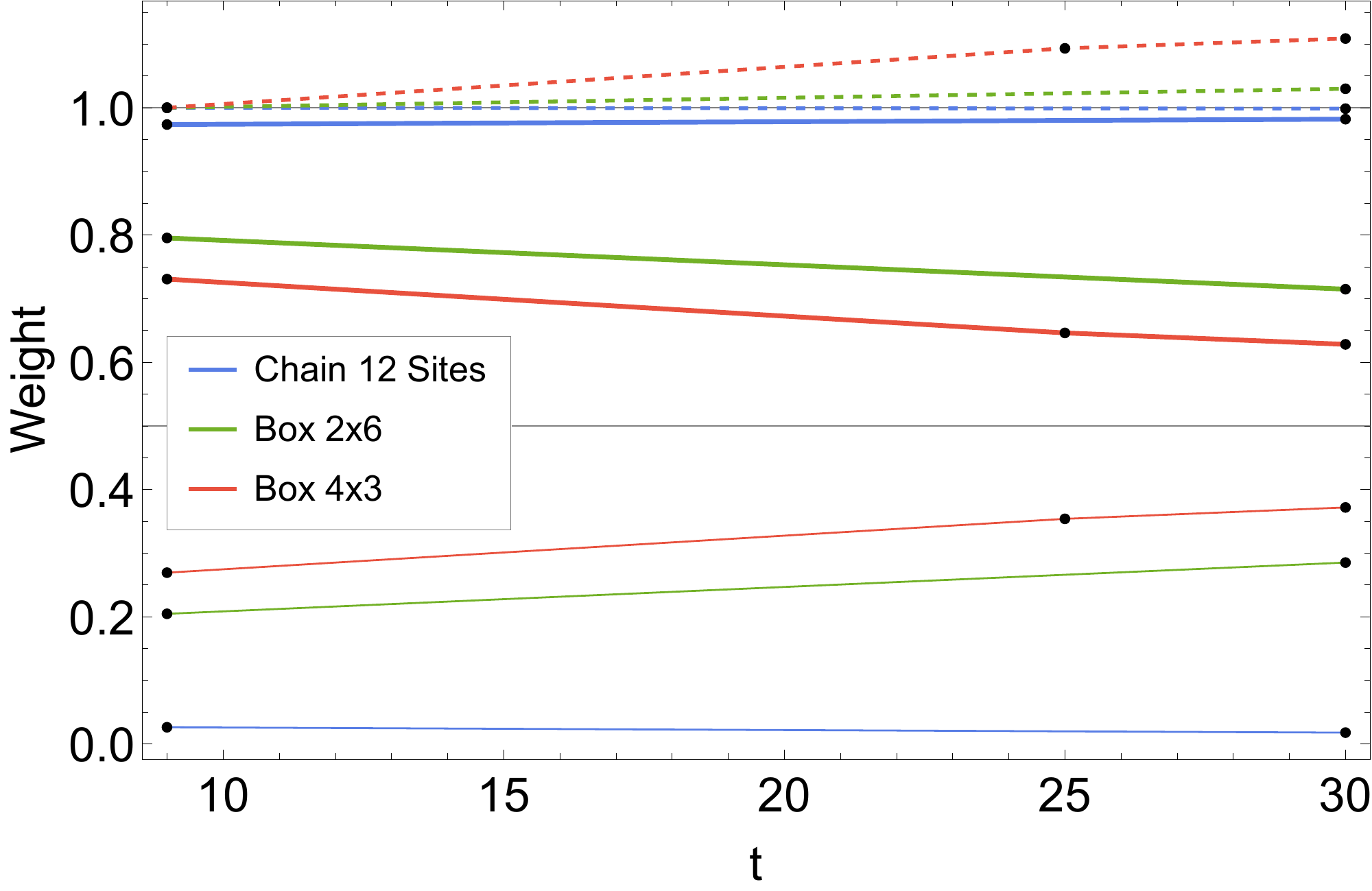}
		 \caption{Spectral weight shifts in the UHB calculated from $A^<(\omega,t)$ shown in Fig.~\ref{fig:spec1} for the three $12$-site geometries. The lines show integrals of $A^<(\omega)$ for a given time $t$ over the following intervals: $[0,4]$ (thick line), $(4,8]$ (thin line), and $[0,8]$ (dashed line). The values are normalized with respect to the integral over $[0,8]$ at $t=9$.
			}
		 \label{fig:uhb1}
\end{figure}

In Fig.~\ref{fig:uhb1} we show the integrated  spectral weight in the thusly defined  upper and lower part of the UHB together with their sum as a function of time after the pulse. We see that in both box geometries there is a spectral weight shift to the lower part of the UHB (the electrons lose the kinetic energy) that is accompanied by a total weight gain in the UHB, which is stronger for the $4\times 3$ geometry. The additional gain originates from impact ionization, which generates  additional doublons. We do not observe such weight shifts in the $12$-site chain. Interestingly, as seen in the $2\times 6$ system, not all kinetic energy lost by the weight shift to smaller frequencies is converted to potential energy (the increase of the total UHB weight is very small). It is in this case mostly compensated by a parallel shift inside the lower Hubbard band (LHB) (cf. the blue and red curves in Fig.~\ref{fig:spec1} b)). This is the first step of thermalization that takes place inside the Hubbard bands, as discussed in detail in ~Ref.\onlinecite{wais2018}.       

Another important difference between the three $12$-site geometries is the behavior of the gap upon pumping energy into the system. All three systems are initially Mott insulators with a gap of similar size. This can be seen in Fig.~\ref{fig:spec1}, where we plot in gray the $A^<(\omega)$ for the equilibrium ground state. After the pulse we see gap filling in both box geometries, but not in the chain. For the chain there remains a clear gap with zero spectral weight between the upper and lower Hubbard band, although it is the chain that initially absorbs the most energy and double occupations (cf. Fig.~\ref{fig:dspec1}). The gap filling is the strongest in the $4\times 3$ box where also impact ionization is the strongest. The photo-induced gap filling has previously been reported in exact diagonalization~\cite{innerberger2020,okamoto2019} and for extended systems in dynamical mean-field theory~\cite{werner2014}, but it is missed by the Boltzmann approach~\cite{wais2018}.

The reason for different behavior of the chain as compared to the  box geometries lies very likely in the difference in dimensionality of these systems. Although the systems we  investigated are very small, the 1D chain and 2D box geometries show qualitatively different behavior. It is not entirely clear whether this difference stems from strong antiferromagnetic (AFM) spin fluctuations in one dimension~\cite{Essler05} or from the fact that the 1D chain hosts fewer nondegenerate eigenstates. For the box geometries we certainly also have AFM fluctuations, but they are weaker than in 1D (see also Sec.~\ref{sec:k_vs_d10}). Also for our system sizes many sites belong to the boundary, which further influences the result (for site-resolved results see Appendix~\ref{app:site_resolved}). We have also searched for signatures of impact ionization in similar systems with periodic boundary conditions, but the parameter scan gave no positive results. In small periodic systems the symmetry is high, which leads to many degenerate energy eigenvalues. In effect there are fewer different energies that the electrons can have after the pulse. Consequently, the range of optimal values of $U$ and $\omega$ needed for impact ionization is strongly reduced.

\subsection{Systems with NN and NNN hopping}
\label{Sec:NNN}

 Since we cannot further increase the dimensionality of the cluster (due to a prohibitively large computational effort), we increased the number of available sites to which electrons can hop by adding a NNN hopping. It increases the connectivity of the system, which we understand here not as the number of NN but as the total number of neighbors $j$ of site $i$ with nonzero $v_{ij}$.

\begin{figure}
	\centering
	\includegraphics[width=\linewidth]{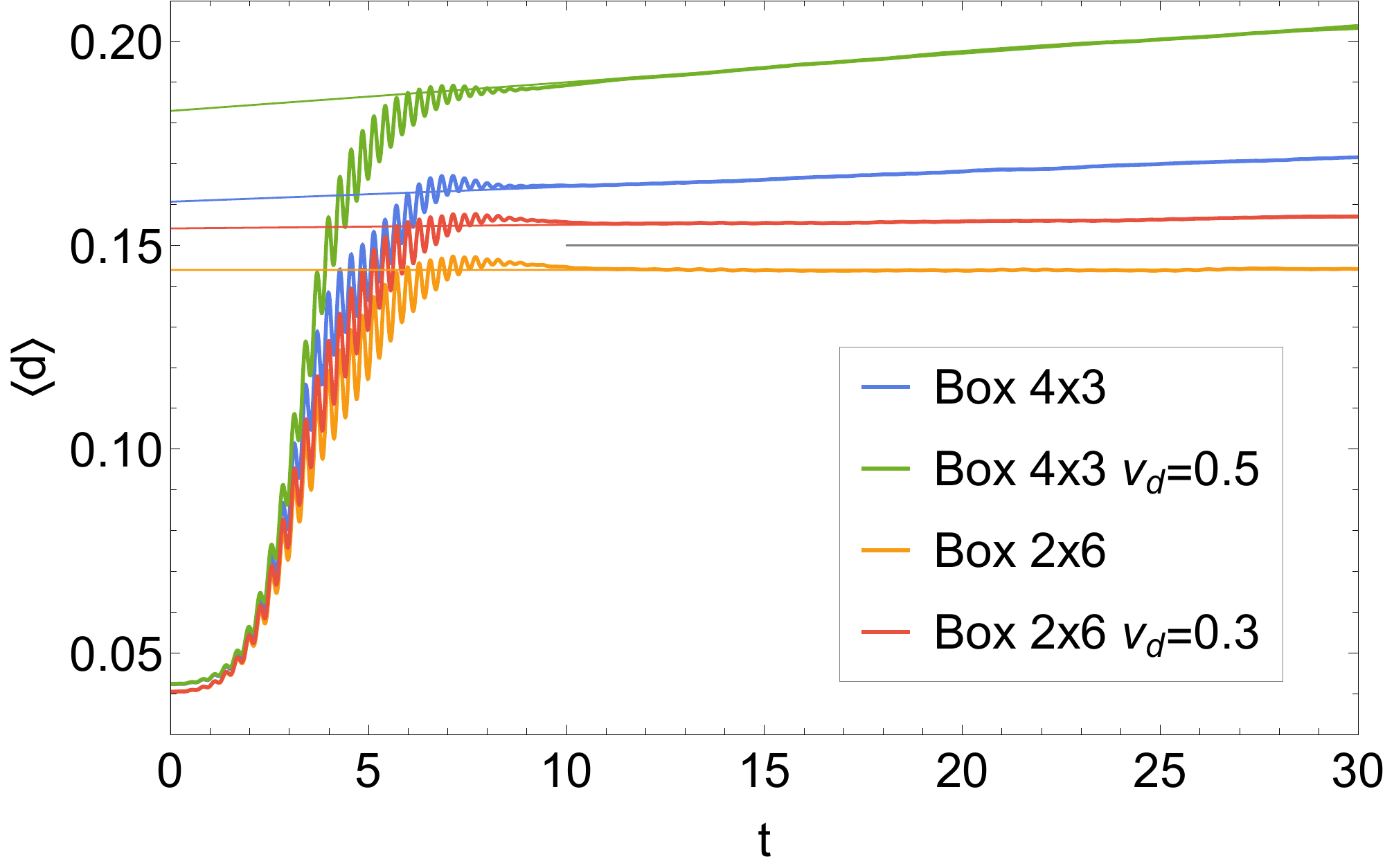}
		 \caption{Average double occupation as a function of time for a $2\times6$ box and a $4\times3$ box with $U=8$,  $\omega_p=11$ and $a=0.8$ (the same parameters as in Fig.~\ref{fig:dspec1}), comparing the boxes without and with NNN hopping $v_d$. The straight lines are linear fits to the data in range $t \in [10,30]$. A horizontal gray line is again added for reference.
			}
		 \label{fig:nnn1}
		 \end{figure}

In Fig.~\ref{fig:nnn1} we present a comparison between the average double occupation for systems without and with NNN hopping. We immediately see that for the same parameters the double occupation after the pulse in the $4\times 3$ box increases significantly steeper with time when NNN hopping is added. That is, NNN hopping boosts impact ionization. There is also improvement for the $2\times 6$ system, but not a very  significant one. Further increasing  the value of $v_d$ does not  enhance  impact ionization for both systems. Let us also note that not only the slope after the photon pulse but also the overall increase of double occupation is bigger when NNN hopping is added (the system can absorb more energy). 

In a significantly smaller system of $8$ sites with $2\times4$ box geometry we also find impact ionization for similar parameters as for the $2\times6$ box (slightly smaller $U$ and $\omega$ seem to give better results, see Appendix~\ref{app:scan} for a parameter scan). In Fig.~\ref{fig:nnn2} we present the average double occupation for different values of the NNN hopping. The bottom (orange) curve is the result with only NN hopping ($v_d=0$). The slope of the increase after the pulse is steeper than in the case of the $2\times 6$ box. It is either due to the fact that we did not find the optimal parameter set for the $2\times 6$ box; or, in the bigger box, AFM spin fluctuations might play a stronger role than in a smaller box, where boundary effects are stronger. The extent of boundary effects can be seen in Appendix~\ref{app:site_resolved}, where we show site-resolved double occupation and spectral function for different systems.

\begin{figure}[t]
		\centering
		 \includegraphics[width=\linewidth]{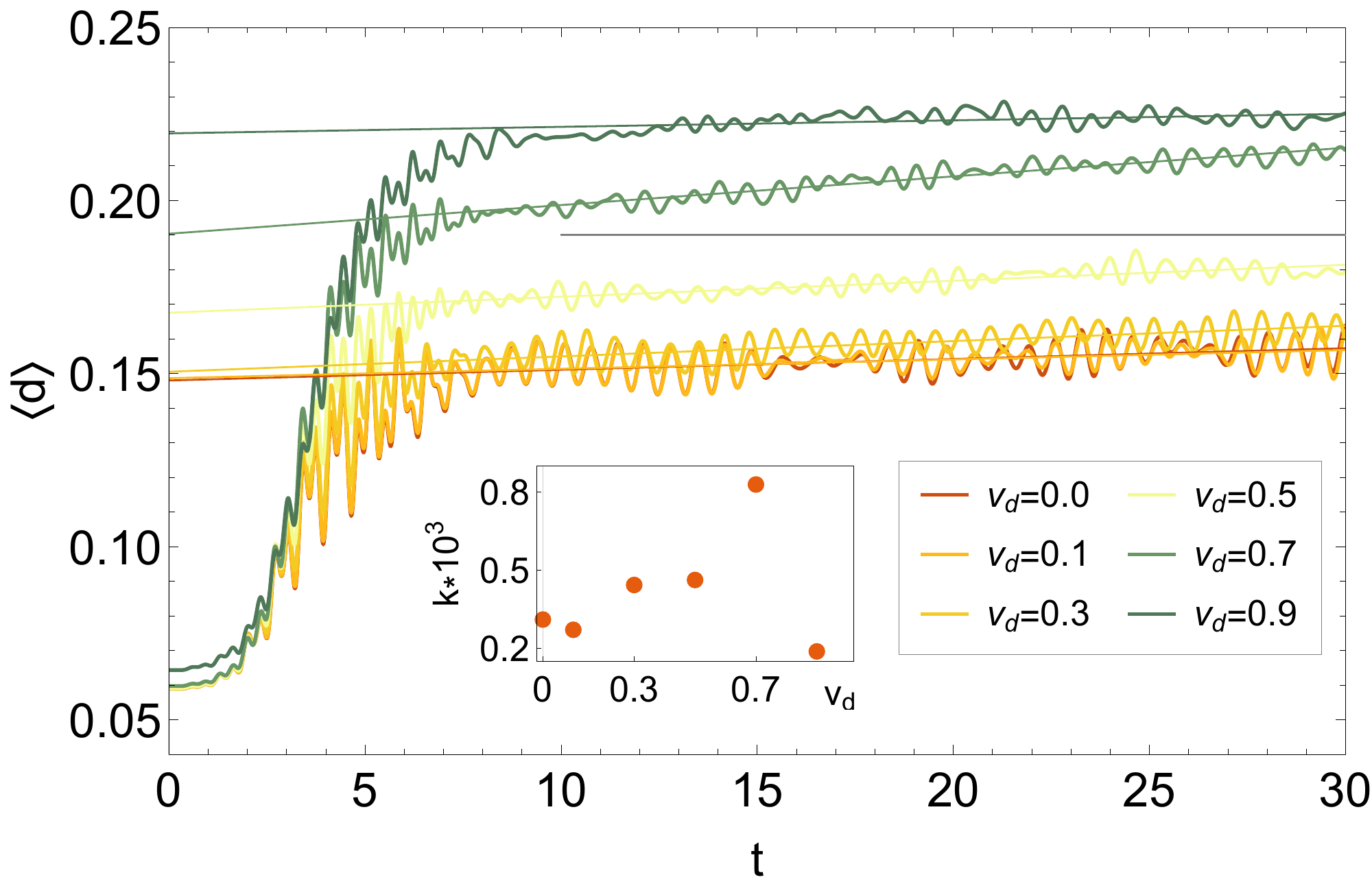}
		 \caption{Average double occupation as a function of time for a $2\times4$ box and different values of the NNN hopping $v_d$. The parameters are: $U=6$,  $\omega_p=9$, and $a=0.8$. The straight lines are linear fits to the data in range $t \in [10,30]$. A horizontal gray line is also added for reference. Inset: Rate of impact ionization $k$ defined as the slope of the linear fits in the  $[10,30]$ time interval as a function of $v_d$.}
		 \label{fig:nnn2}
\end{figure}

To quantify the enhancement of impact ionization upon increasing the NNN hopping we define the rate of impact ionization $k$ as the slope of the linear fit to the data points in the time interval $[10,30]$ \footnote{The measure of the secondary rise of double occupation could also be e.g. the ratio of the finally obtained double occupancies (at $t=30$) to the once present directly after the pulse (at $t=10$). But since the double occupation oscillates in time, it would require some averaging over time. We found the linear coefficient of the fit between $t=10$ and $t=30$ to be more reliable and not sensitive to oscillations.}. In the inset of Fig.~\ref{fig:nnn2} we show how $k$, i.e., the rate of impact ionization, depends on the strength of NNN hopping $v_d$. This dependence is not monotonous and reaches a maximum at $v_d\approx 0.7$. As we show later, in Sec.~\ref{Sec:Disc}, when the overall average double occupation reaches $\approx 0.2$, the rate of impact ionization gets lower.

\subsection{Triangular lattice with $10$ sites}
\label{Sec:Tri}
Another qualitatively different geometry is obtained by taking a fragment of a triangular lattice as depicted in Fig.~\ref{fig:tri10}. This system  has only NN hopping, but the EM vector potential influences hoppings in different directions differently, as explained in Sec.~\ref{sec:peierls} and also illustrated in  Fig.~\ref{fig:tri10}. Due to frustration AFM fluctuations are suppressed (see also Sec.~\ref{sec:k_vs_d10}). Among the systems studied, this system has  (together with the $4\times 3$ box with NNN hopping)  the largest connectivity. 

\begin{figure}
		 \includegraphics[width=\linewidth]{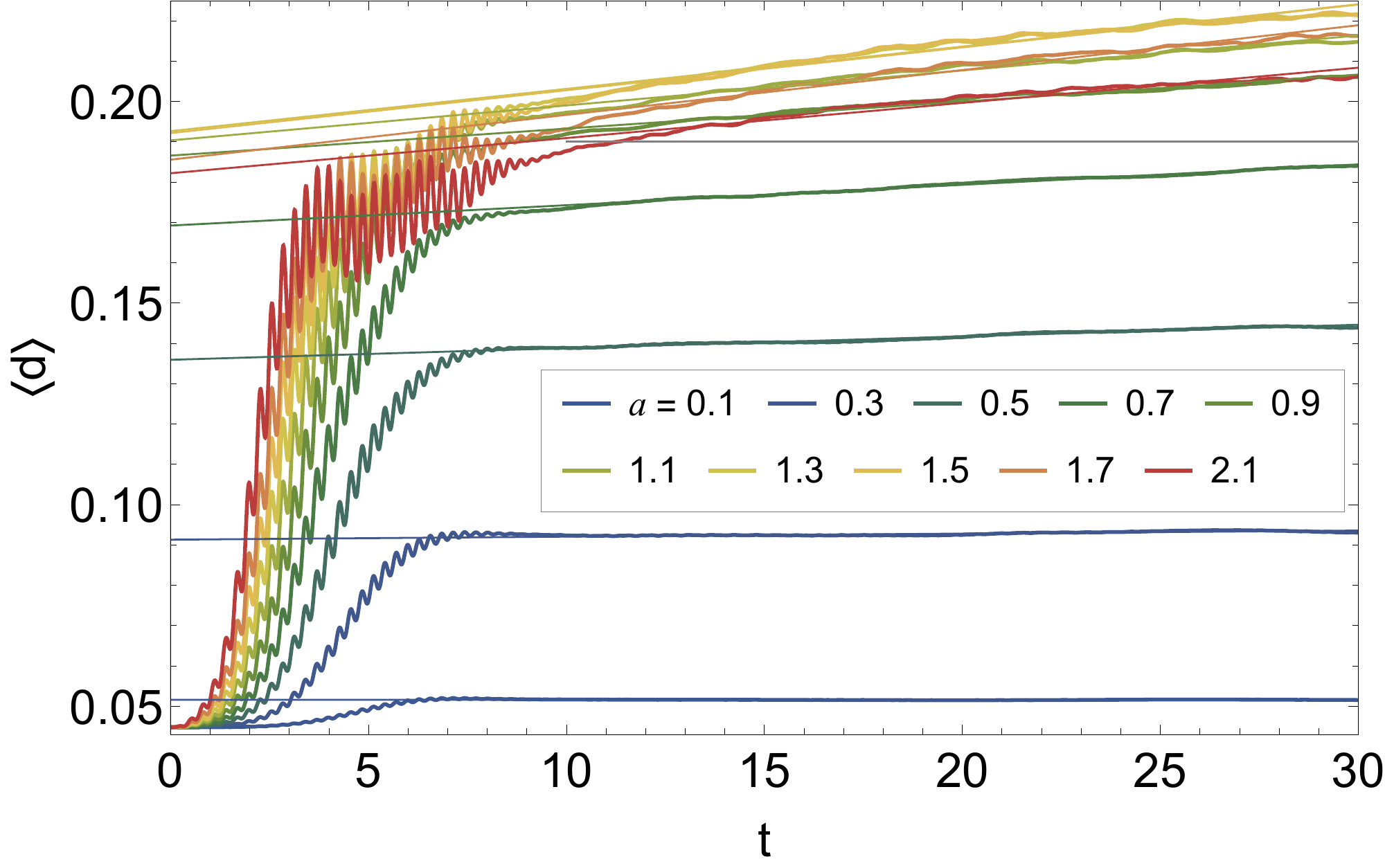}
		 \caption{Average double occupation as a function of time for the $10$-site cluster (tri-$10$) depicted in Fig.~\ref{fig:tri10} and different values of the pulse strength $a$. The parameters are: $U=8$,  $\omega_p=11$. The straight lines are linear fits to the data in range $t \in [10,30]$. A horizontal gray line is also added for reference.
		}
		 \label{fig:dtri}
\end{figure}

\begin{figure}
	\centering
		 \includegraphics[width=\linewidth]{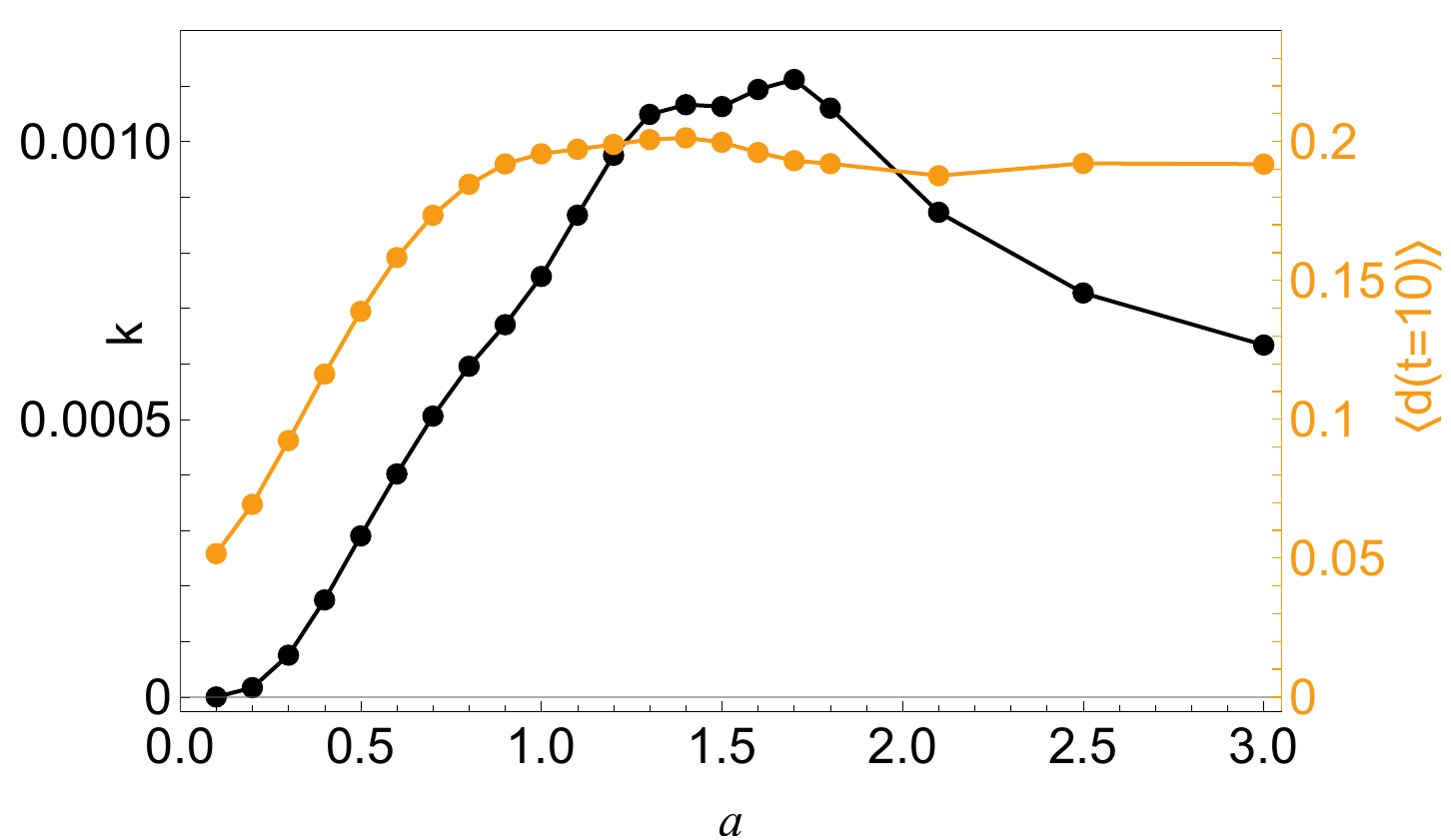}
		 \caption{Rate of impact ionization $k$ obtained from the fits to data in Fig.~\ref{fig:dtri} (left vertical axis, black) and the average double occupation at $t=10$ (right vertical axis, orange) as a function of the pulse strength $a$ for the 10-site triangular lattice cluster (tri-$10$). Same parameters as in Fig.~\ref{fig:dtri}.
}
		 \label{fig:tri1}
\end{figure}

In Fig.~\ref{fig:dtri} we show the average double occupation for the  $10$-site triangular lattice fragment (tri-$10$ cluster) for different values of the pulse strength $a$. In this geometry we find the impact ionization is the most pronounced and noticeable already for small pulse strengths, when the overall increase of double occupation is relatively small. As already seen in the inset of Fig.~\ref{fig:nnn2} for the $2\times 4$ box, the rate of impact ionization reaches its maximum when the overall double occupation is around $0.2$ and then gets lower. This is better visible in  Fig.~\ref{fig:tri1}, where we show the rate of impact ionization $k$ and the double occupation directly after the pulse (at $t=10$) as a function of the pulse strength $a$. While the double occupation after the pulse increases approximately linearly with the pulse strength $a$ for moderate values of $a$, the rate of impact ionization, quite counter-intuitively, only grow sub-linearly for small values of $a$.

\subsection{Systems with disorder}
\label{Sec:disorder}

In semiconductor solar cells randomly distributed impurities are a source of large inefficiencies due to additional in-gap states. The generated electrons and holes may get trapped at these impurities, decreasing the overall electrical current and energy harvested in the solar cell. In transition metal oxides a similar defect trapping may be expected for the prevalent defects: oxygen vacancies.  We are unable here to address this trapping of the generated current as it requires much larger systems than we can study with exact diagonalization. For small clusters and without the transport effect, we observe a quite interesting, opposite effect:  introducing disorder lowers the symmetry of the systems and opens more possibilities for energy match for impact ionization\footnote{As we already mentioned at the end of Sec.~\ref{sec:nn_only}, high symmetry of the system leads to fewer eigenvalues of the Hamiltonian and this is not favorable for impact ionization} (see also Appendix~\ref{app:degeneracies}). Impact ionization is strongly enhanced by disorder, increasing the efficiency of solar energy conversion.

\begin{figure}
		\centering
		 \includegraphics[width=\linewidth]{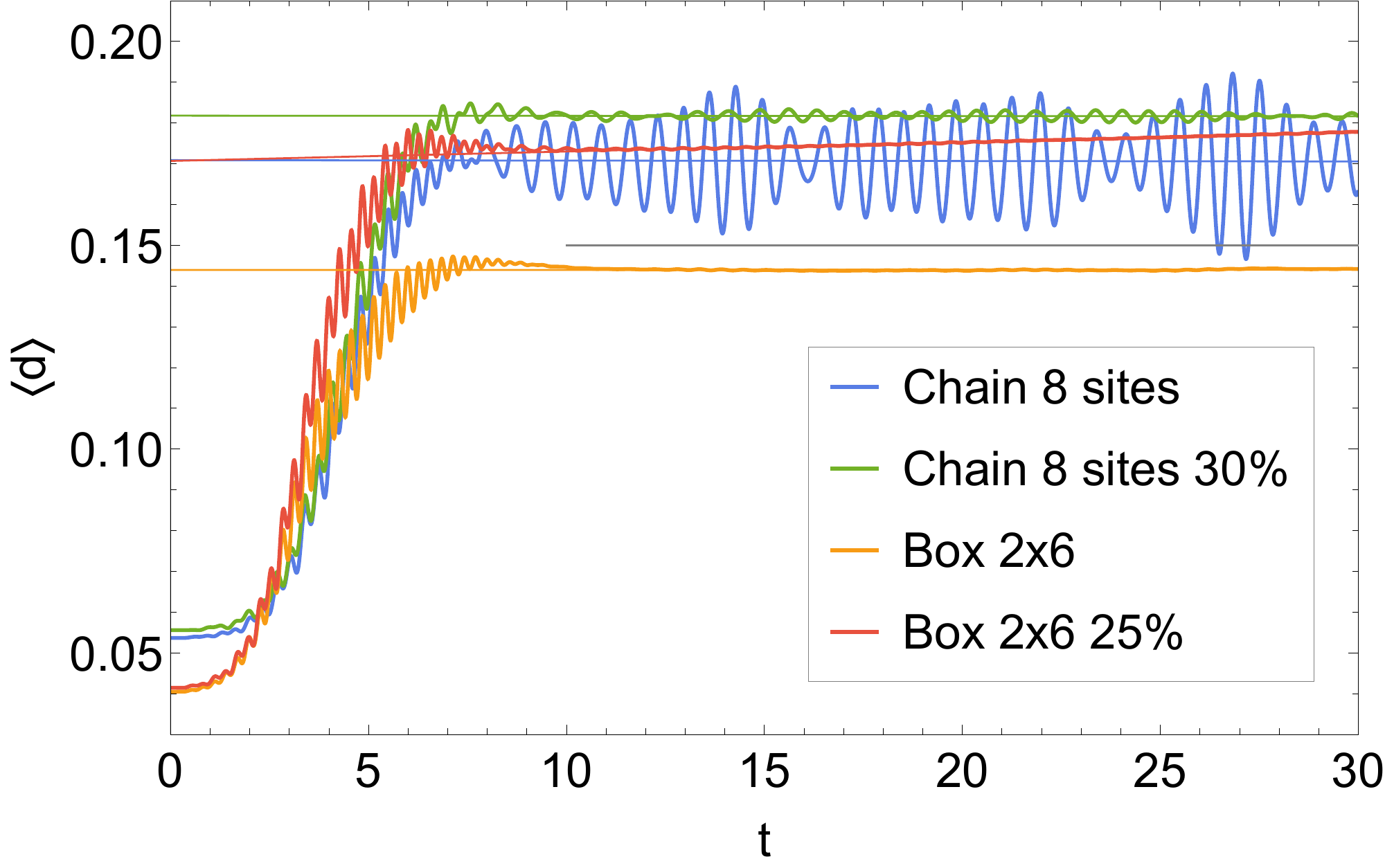}
		 \caption{Comparison of the average double occupation for the  $8$-site chain and the $2\times6$ box for cases with and without disorder. The data points for disordered systems are averages over $N_{seeds}=31$ disorder realizations. The parameters are: $U=6$,  $\omega_p=9$, and $a=0.6$ for the $8$-site chain and $U=8$,  $\omega_p=11$, $a=0.8$ for the $2\times6$ box.
The disorder strength is  $\Delta/U=30\%$ and 25\%, respectively.
 The straight lines are linear fits to the data in range $t \in [10,30]$. A horizontal gray line is also added for reference.
}
		 \label{fig:dis1}
\end{figure}

We introduce site disorder by changing the values of the on-site potentials as described in Eqs.~\eqref{eq:disorder}-\eqref{eq:onsite} (box disorder). The results we present in the following are averaged (arithmetically) over disorder realizations. In case of the chain geometry we still do not observe impact ionization (see Fig.~\ref{fig:dis1}), but for the $2\times6$ box there is a slightly stronger rise of double occupation when we add $25\%$ disorder (the percentage given in all plots is the ratio $\Delta/U\cdot 100\%$). 

\begin{figure}
		\centering
		 \includegraphics[width=\linewidth]{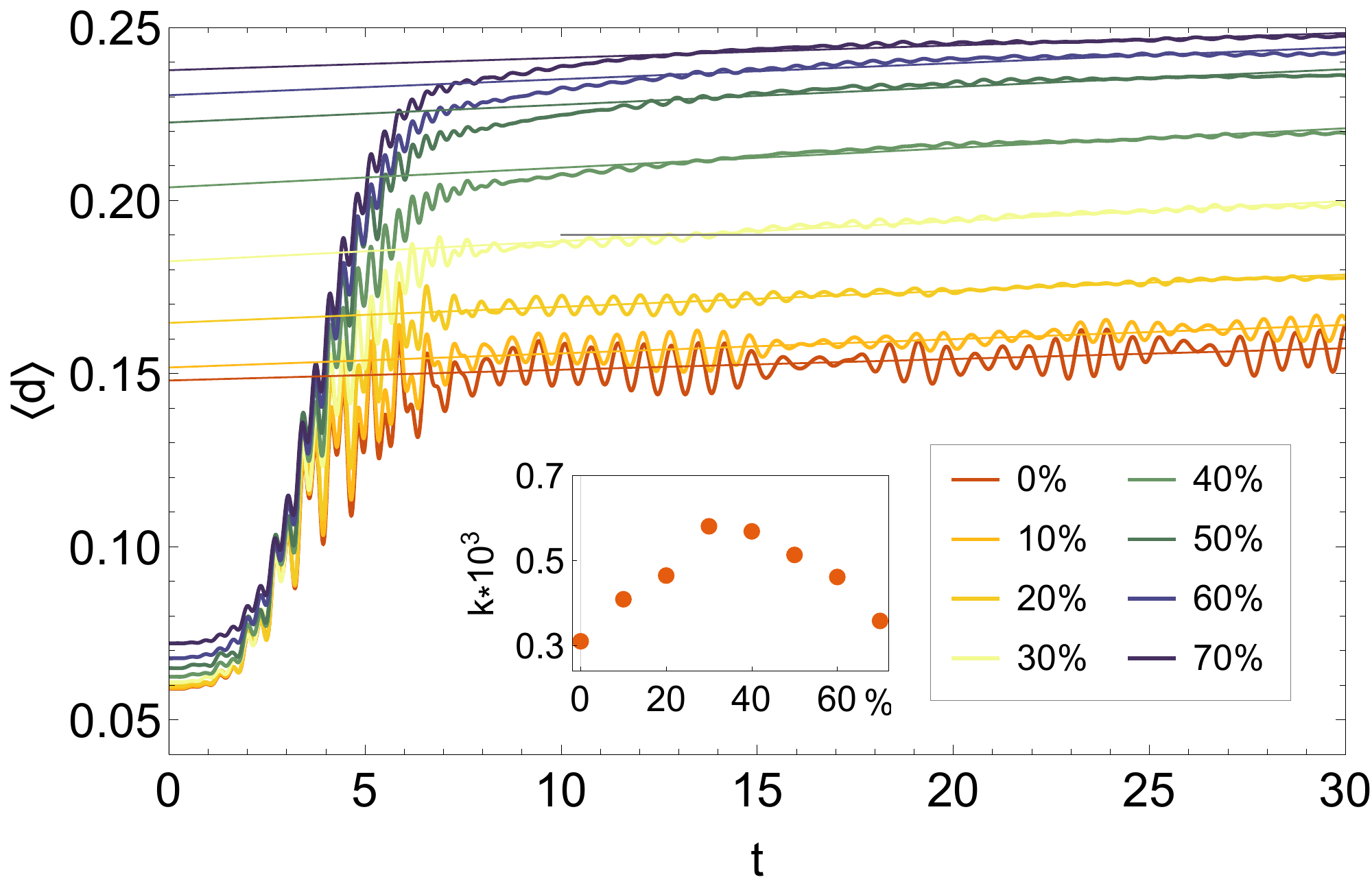}
		 \caption{Average double occupation for the $2\times4$ box for different disorder strengths $\Delta/U$ (in \%). The data points  are averages over $N_{seeds}=31$ disorder realizations. The parameters are: $U=6$,  $\omega_p=9$, and $a=0.8$. The straight lines are linear fits to the data in range $t \in [10,30]$. A horizontal gray line is also added for reference. Inset: Rate of impact ionization $k$ as a function of disorder strength $\Delta/U$  (in \%).}
		 \label{fig:dis2}
	
\end{figure}

For the $2\times4$ system, which is computationally less demanding, we show in  Fig.~\ref{fig:dis2} the time dependent double occupation for several disorder strengths. Adding disorder leads to a strong increase of the rate of impact ionization by up to a factor of two, with a maximum again at the point when the double occupation after the pulse is $\approx0.2$.

Please note, that our random potential is a highly idealized model. In real materials there may be e.g.\ oxygen vacancies in transition metal oxides that all have the same potential strength, which rather corresponds to a binary disorder distribution. However, there are further adatoms, other impurities, and lattice mismatches so that a binary disorder distribution is similarly idealized.

\subsection{General considerations}
\label{sec:k_vs_d10}
\label{Sec:Disc}

The different ways of modifying the Hubbard model, changing the geometry, adding the NNN hopping or disorder, seem all to influence the system in the direction of stronger impact ionization. Except for the chain geometry, where we did not observe impact ionization at all ($k<10^{-6}$ for chains of up to $14$ sites).

\begin{figure}
		\centering
		 \includegraphics[width=\linewidth]{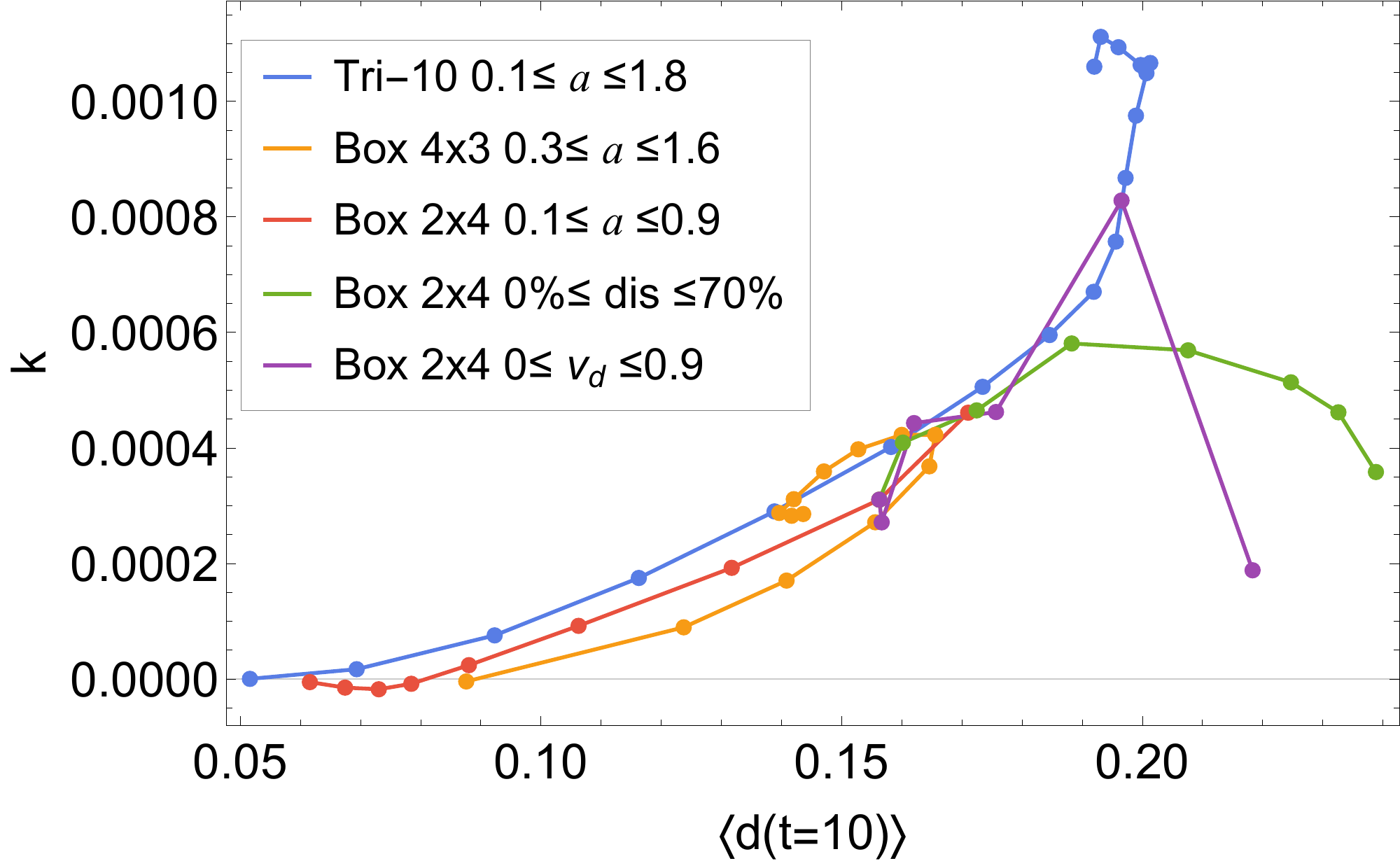}
		 \caption{Rate of impact ionization $k$ vs. mean double occupation at time $t=10$, $\langle d(t=10)\rangle$, for different scans of parameters (see legend box; $a$: pulse strength, ${\rm dis}=\Delta/U$: disorder strength, $v_d$: NNN hopping strength) for the $2\times4$ and $4\times3$ boxes and the triangular lattice with $10$ sites (tri-$10$ cluster). Parameters for the $2\times4$ box: $U=6$, $\omega_p=9$, $a=0.8$ (or as in the legend); for the $4\times3$ box and 
	tri-$10$ cluster: $U=8$, $\omega_p=11$ and $a$ as in the legend.}
		 \label{fig:k_vs_d10}
\end{figure}

In Fig.~\ref{fig:k_vs_d10} we show the rate of impact ionization for different system as a function of double occupation directly after the pulse, i.e., at $t=10$. Different double occupations are obtained by changing either the pulse strength $a$, the NNN hopping $v_d$, or the disorder strength $\Delta/U$. This way, we can plot the results from the previous figures and additional data in a single figure.

\begin{figure}
		 \includegraphics[width=1.05\linewidth]{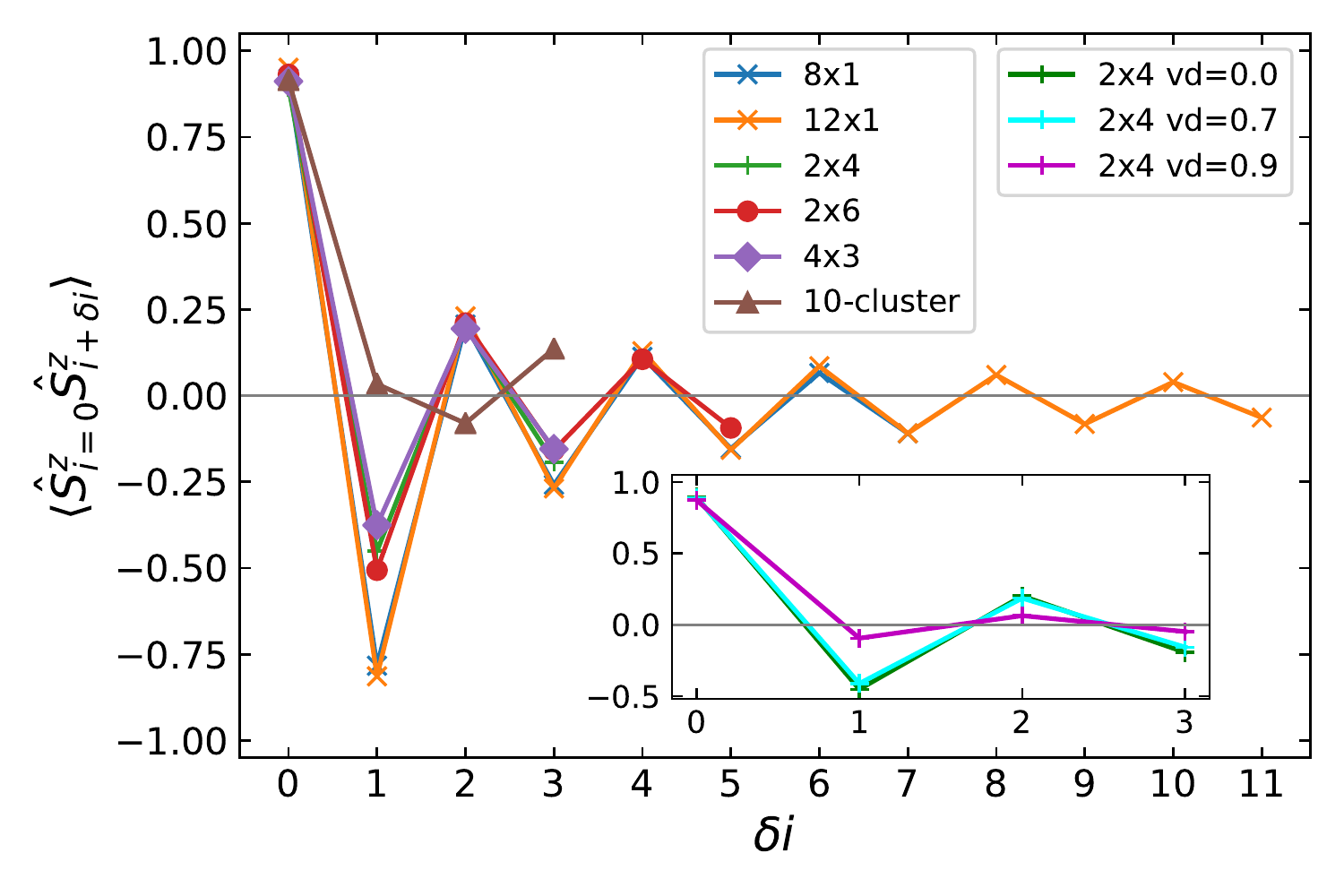}
		 	 \caption{The static spin-spin correlation function $\langle \hat{S}^z_i(t=0)\hat{S}^z_{i+\delta i}(t'=0) \rangle$  for different systems as a function of the distance $\delta i$ from site $i=0$ in the horizontal direction. The site $i=0$ is the leftmost site for chains ($8\times1$ and $12\times1$) and double chains ($2\times4$ and $2\times 6$ and the leftmost middle site for $4\times 3$ and tri-$10$ cluster. The same correlation function for the $2x4$ box with different NNN hoppings $v_d$ is shown in the inset.}
		 \label{fig:spin_spin}		
	\end{figure}
	
In all cases the rate $k$ increases with increasing $\langle d(t=10)\rangle$ and reaches a maximum when $\langle d(t=10)\rangle \approx 0.2$. This somewhat universal behavior indicates that the rate of impact ionization depends in the first place on the number of initially generated double occupations (doublons). Since the number of the latter is limited for a finite number of electrons, we see a clear maximum.

\begin{figure}
		 \includegraphics[width=\linewidth]{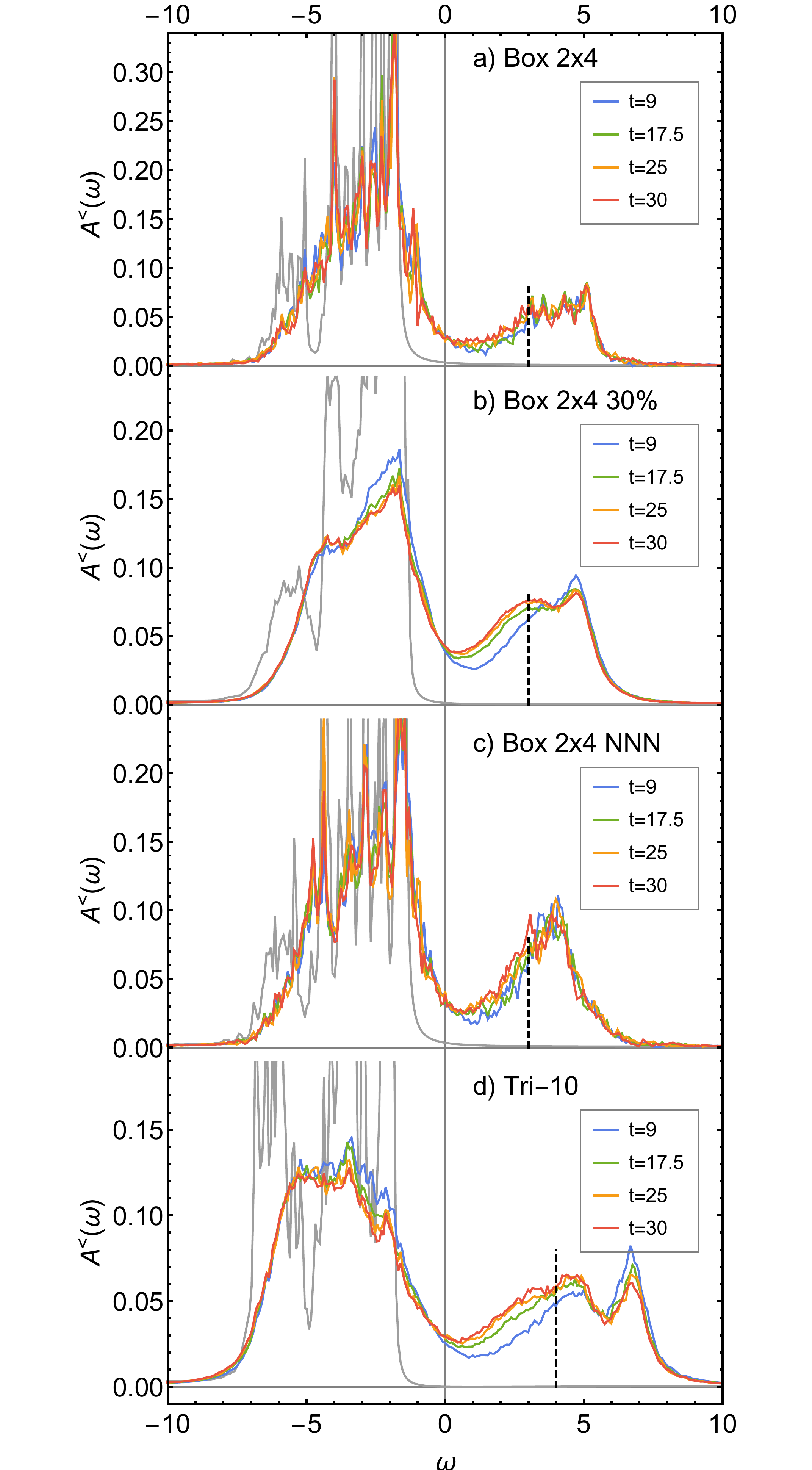}
		 	 \caption{$A^<(\omega,t)$  at different times $t$ for the a)  $2\times4$ box,  b)  $2\times 4$ box with $30\%$ disorder,  c) $2\times4$ box with NNN hopping $v_d=0.3$, and d) $10$-site  cluster (tri-$10$). The parameters are $U=6$, $\omega_p=9$, $a=0.8$ for a)-c) and $U=8$, $\omega_p=11$, $a=0.8$ for d).  The dashed line indicates the separation into lower and upper part of the UHB used in Fig. \ref{fig:uhb2}. The gray curve is $A^<(\omega)$ in the ground state. The results are broadened with $\epsilon$=0.04.}
		 \label{fig:spec2}		
	\end{figure}

For smaller values of $\langle d(t=10)\rangle$ the behavior of $k$ is more differentiated between the systems. For both box geometries there seems to be a certain threshold of $\langle d(t=10)\rangle$  below which we do not observe impact ionization. This is not the case for the geometrically frustrated triangular geometry where AFM fluctuations are suppressed.

As an illustration of the suppression of AFM fluctuations in the frustrated geometry, we present in Fig.~\ref{fig:spin_spin} the static spin-spin correlation function $\langle \hat{S}^z_i\hat{S}^z_{i+\delta i} \rangle$  (with $\hat{S}^z_i = n_{i\uparrow}-n_{i\downarrow}$) for different systems as a function of the distance $\delta i$ from site $i$ in the horizontal direction.  We choose the leftmost site $i$ for chains and double chains and the leftmost middle site for $4\times 3$ and tri-$10$ geometries in order to show the longest possible distance~\footnote{The overall picture does not change qualitatively if we pick a different site $i$, but the distances we can show become shorter}. For both chains shown ($8\times1$ and $12\times1$) the AFM correlations are the strongest. With changing the geometry to double chains, there is a significant drop in strength, especially for the first neighbor. This drop is slightly stronger for the $2\times 4$ box than for the $2\times6$ box. For the $4\times3$ box we see a still stronger decrease and for the fully frustrated tri-$10$ cluster we do not see AFM correlations at all. 

The spin-spin correlation function for different NNN hoppings in the $2\times4$ cluster, shown in the inset of Fig.~\ref{fig:spin_spin}, also shows suppression of AFM correlations but only when the NNN hopping is strong. This does not correlate with the increased impact ionization already for $v_d=0.3$ in this system. However, even for small values of $v_d$ the number of degenerate eigenstates of the system is decreased (see Appenidx~\ref{app:degeneracies}) and there are more possibilities for an energy match in the process of generating additional electron-hole pairs. This is the case also for systems with disorder.        

\begin{figure}
		 \includegraphics[width=\linewidth]{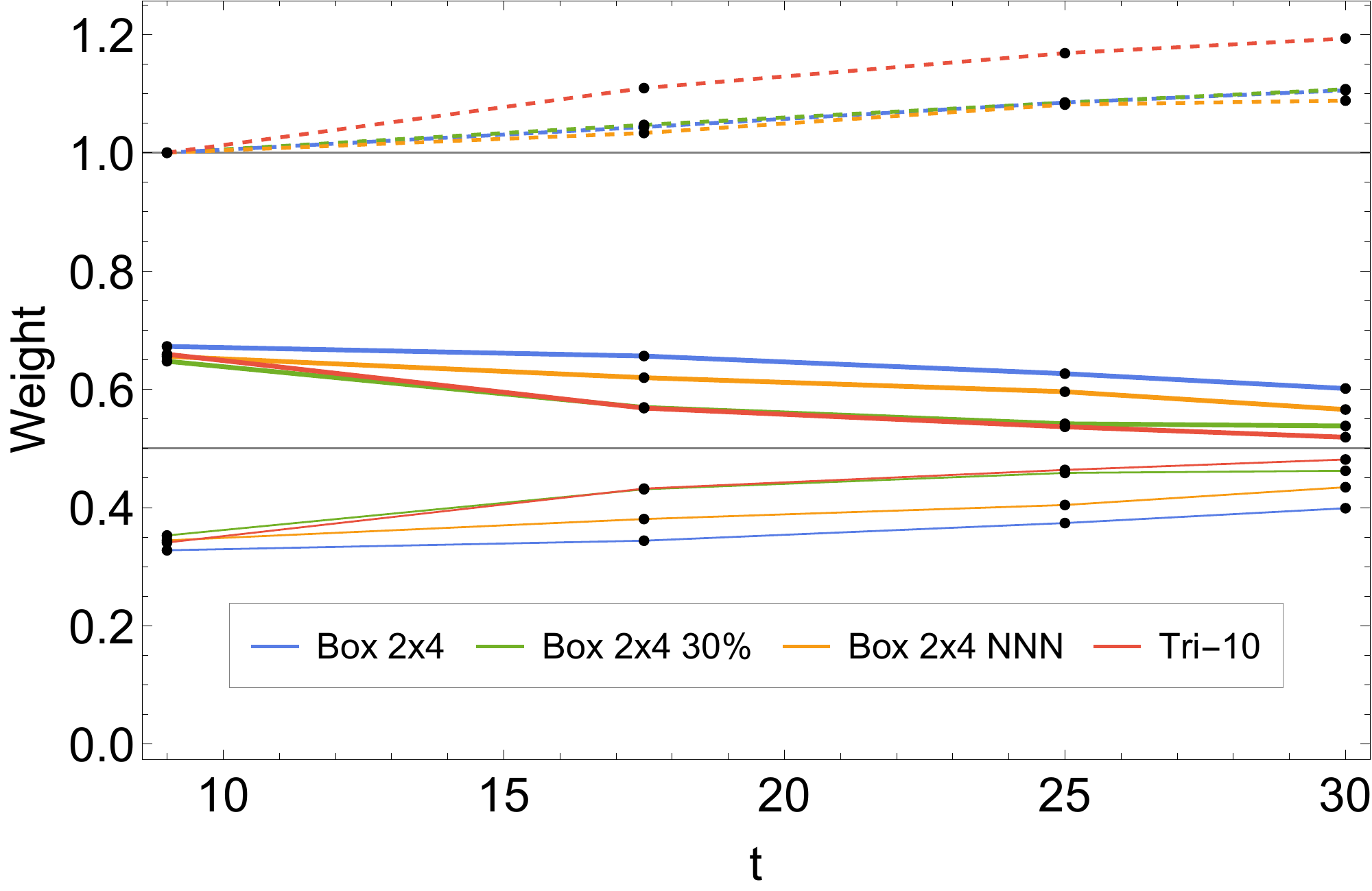}
		 		 \caption{Spectral weight shifts in the UHB calculated from $A^<(\omega,t)$ shown in Fig.~\ref{fig:spec2} for the four cases a)-d). The lines show integrals of $A^<(\omega)$ for a given time $t$ over the following intervals for $2\times4$ box: $[0,3]$ (thick line), $(3,6]$ (thin line), and $[0,6]$ (dashed line); and for the tri-$10$-cluster: 
		 		 $[0,4]$ (thick line), $(4,8]$ (thin line), and $[0,8]$ (dashed line). The values are normalized with respect to the integral over $[0,6]$ (box) or $[0,8]$ (tri-$10$ cluster) at $t=9$.}
		 \label{fig:uhb2}
\end{figure}

The spectral weight shifts inside and between Hubbard bands that we described in Sec.~\ref{sec:nn_only} for $12$-site systems are also present in the smaller systems we studied. In Fig.~\ref{fig:spec2} we show $A^<(\omega,t)$ for several times after the pulse. In all cases we see the spectral weight shift from the upper part of the UHB to its lower part and the reduction of spectral weight in the LHB. Analogously to Fig.~\ref{fig:uhb1}, we also plot for the data in Fig.~\ref{fig:spec2} the integrated spectral weight as a function of time for regimes that we define as lower and upper part of the UHB. The overall behavior of the spectral weight is similar in all four cases and also analogous to the $4\times3$ system.


\section{Summary}
\label{Sec:Summary}
We have studied the time-evolution of the double {occupation} and the (lesser) spectral function for small Hubbard clusters during and after an electric field pulse which models the impact of a photon. A particular focus of our work is the study of impact ionization which creates more than a single electron-hole pair per photon. The additional electron-hole pairs are generated after the photon pulse. This genuine correlation effect bears some potential for applications as it can boost the efficiency of solar cells. More electron-hole pairs mean a larger current and more electrical energy.

We  study  a number of different geometries and parameter ranges, and find, as in Ref.\ \onlinecite{Maislinger2020}, impact ionization for clusters with box and triangular geometries, but not for strictly one-dimensional chains. It can be strongly enhanced when including next-nearest neighbor hoppings, the {geometrical} frustration and larger connectivity of a triangular lattice, or disorder. All of these variations of the Hamiltonian have in common that they both, increase the number of nondegenerate eigenstates and suppress antiferromagnetic fluctuations. The latter may be unfavorable for impact ionization since the first electron-hole pair created by the photon can transfer its excess energy to rearrange (disorder) the spin background instead of creating a second electron-hole pair. The larger number of nondegenerate eigenstates might be more relevant when studying impact ionization for a small cluster, but is possibly less crucial for extended systems.  Our results hence still call for a complementary study in the thermodynamic limit. Irrespective of this caveat,  we have demonstrated  that one can actually study impact ionization for small clusters, and we identified some recipes to enhance it.

Even if some of our findings are modified  for thick films or bulk-like Mott insulators they nonetheless have experimental relevance.  Impact ionization has also been observed in quantum dots\cite{Franceschetti06} possibly bridged to a reservoir through small (hydrocarbon) molecules\cite{Wang13,Wang17}. Indeed, Hubbard-type models are suitable for describing the conjugated $\pi$-bonds in such molecules \cite{Pople53,Pariser53a} as well as simple quantum dots. While our study is  on the most basic model level, the effect that disorder enhances impact ionization  might be exploited here, twisting the molecule or the shape of the quantum dot.

\begin{acknowledgements}
 We thank H.-G. Evertz, F. Maislinger, O. Koch, W. Auzinger and H. Hofst\"atter for many fruitful discussions. The authors acknowledge support from the Austrian Science Fund (FWF):  AK, PW, CW and KH through grant P 30819 and MI through grant W 1245 and the SFB „Taming Complexity in PDE systems“ (FWF grant F 65). Numerical computations were performed in part on the Vienna Scientific Cluster (VSC).
 \end{acknowledgements}

\appendix

\section{Site-resolved double occupation and spectral functions}
\label{app:site_resolved}
\begin{figure}[b]
	\centering
		 \includegraphics[width=0.5\textwidth]{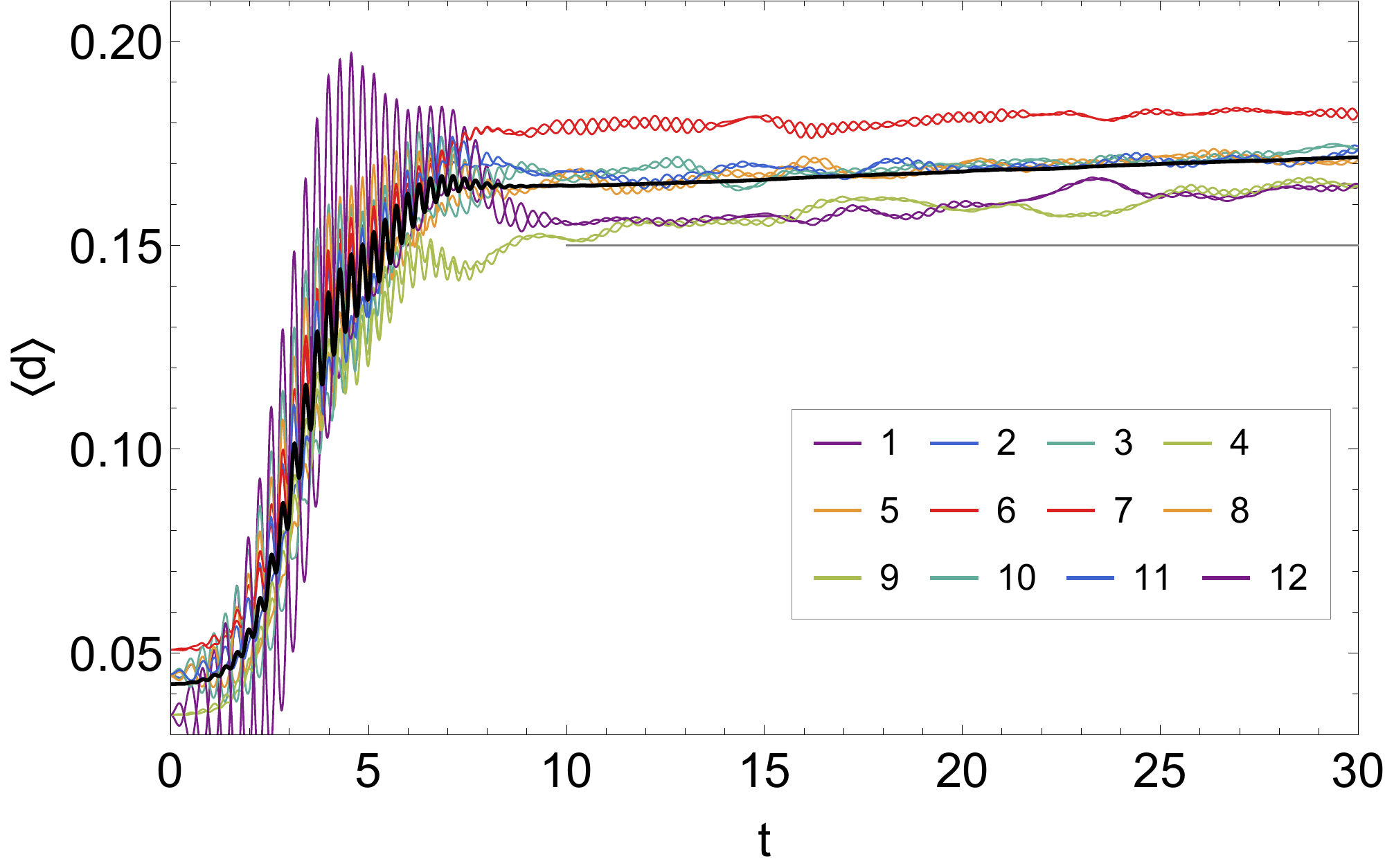}
		 \caption{Site-resolved double occupation for the $4\times3$ box for the same parameters as in Fig.~\ref{fig:dspec1}. Different colors correspond to different sites arranged as shown in the legend. The black curve shows the average over all sites. The horizontal grey line serves only as a guide to the eye. 			}
		 \label{fig:b43_d_sites}
\end{figure}

\begin{figure*}
	\centering
		 \includegraphics[width=0.48\textwidth]{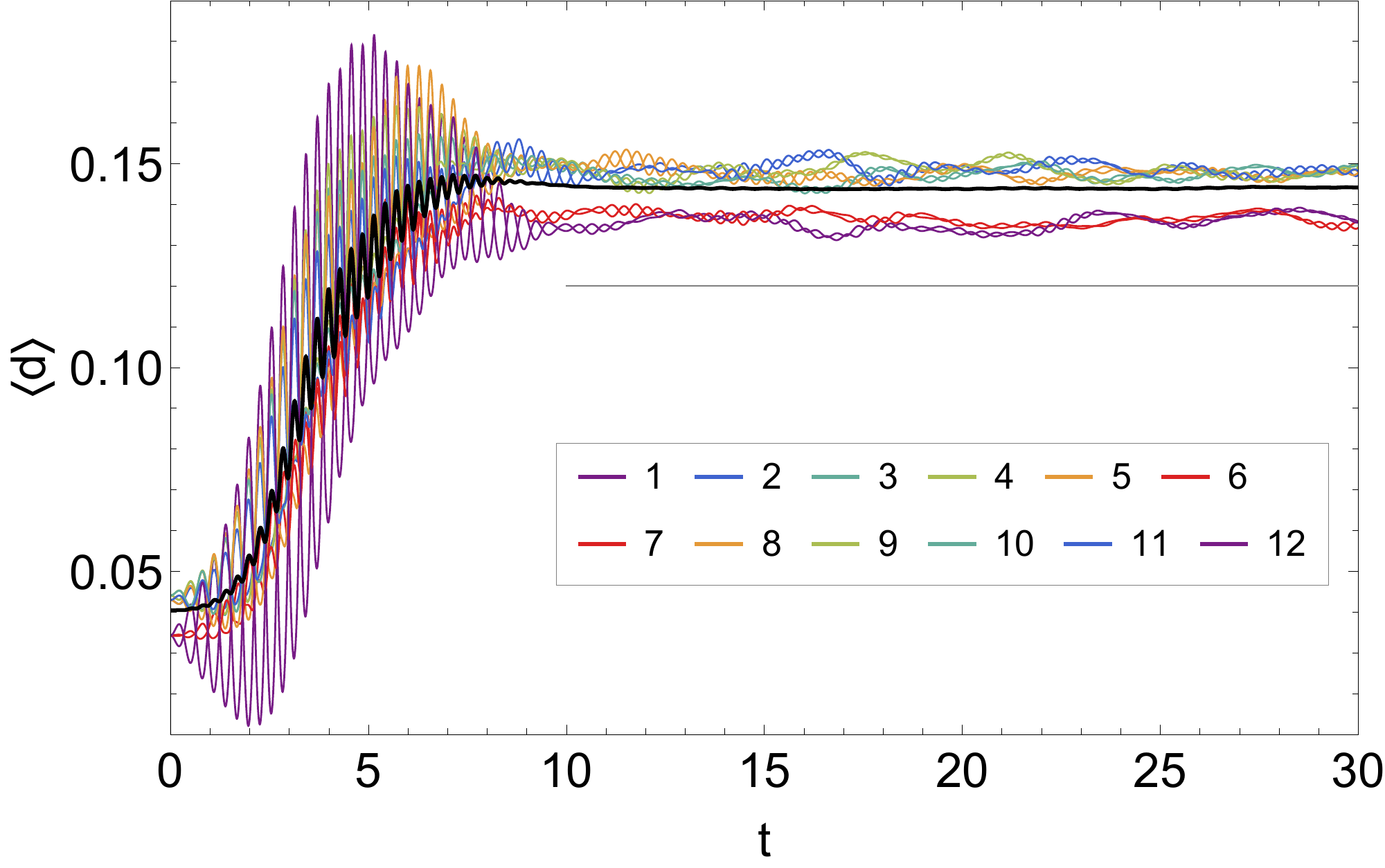}
		 		 \includegraphics[width=0.48\textwidth]{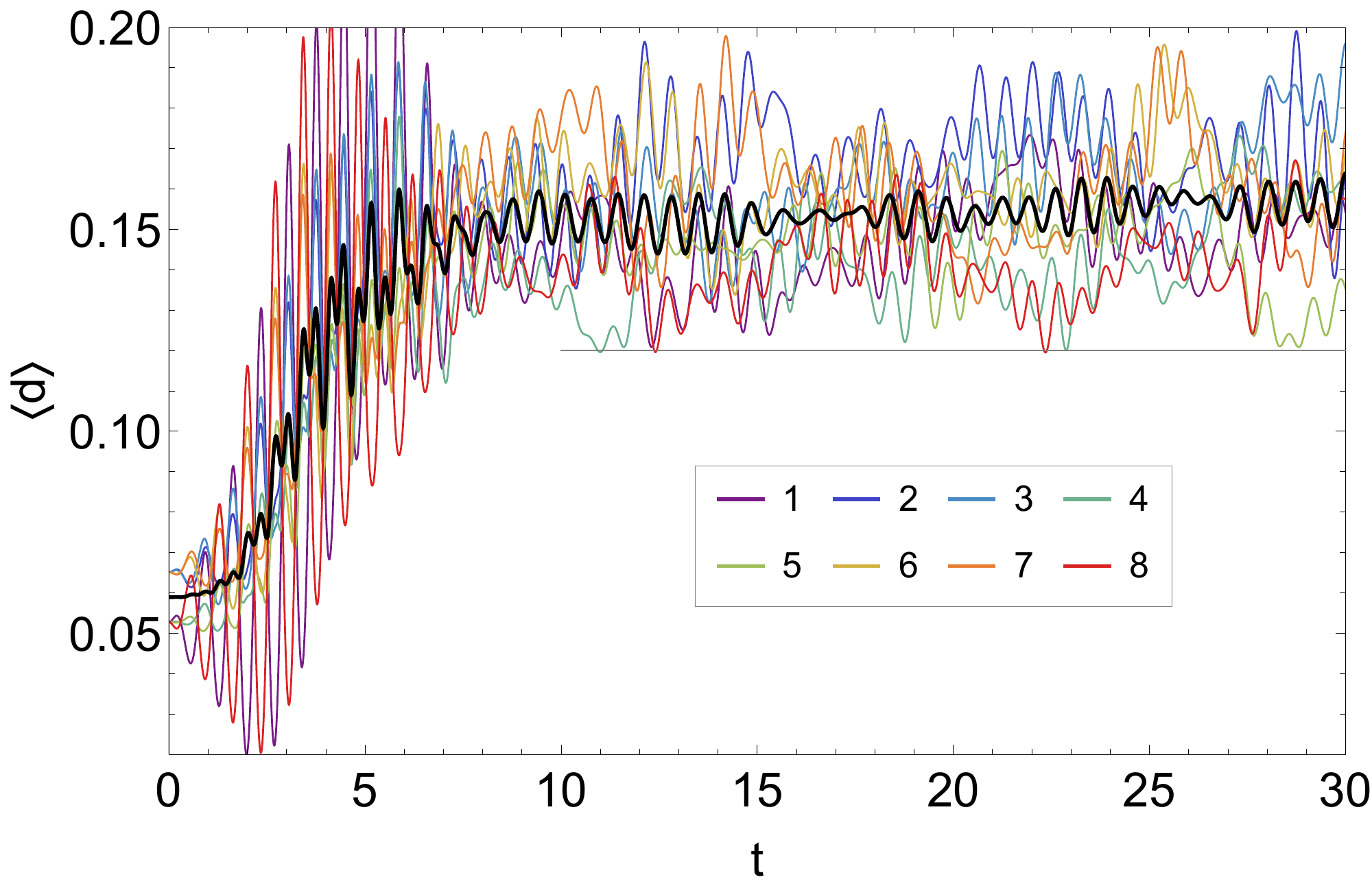}
		 \caption{Site-resolved double occupation for the $2\times6$ box (left)  for the same parameters as in Fig.~\ref{fig:dspec1} and for the $2\times4$ box (right) for the same parameters as in Fig.~\ref{fig:nnn2}  with $v_d=0$. Different colors correspond to different sites arranged as shown in the legend. The black curve shows the average over all sites. The horizontal grey line serves only as a guide to the eye.
			}
		 \label{fig:b26_d_sites}
\end{figure*}
\begin{figure*}
	\centering
		 \includegraphics[width=\textwidth]{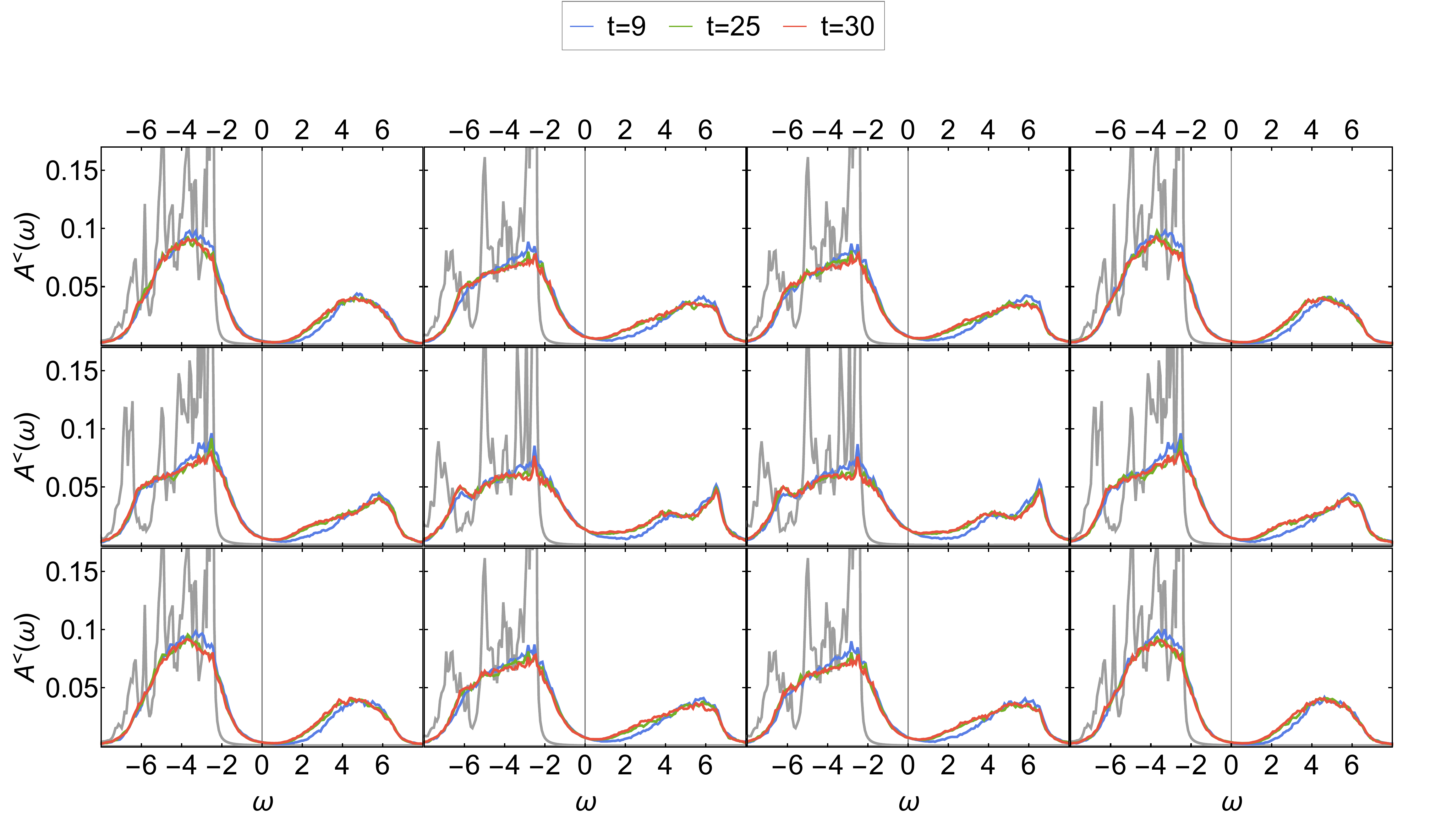}
		 \caption{Site-resolved spectral function $A^<(\omega)$ for the $4\times 3$ box for the same parameters as in Fig.~\ref{fig:dspec1}. The arrangement of the plots corresponds to the arrangement of sites. Different colors correspond to different times after the pulse as shown in the legend at the top. Grey line corresponds to  $A^<(\omega)$ for $t=0$.
			}
		 \label{fig:b43_spec_onsite}
\end{figure*}
\begin{figure*}
	\centering
		 \includegraphics[width=\textwidth]{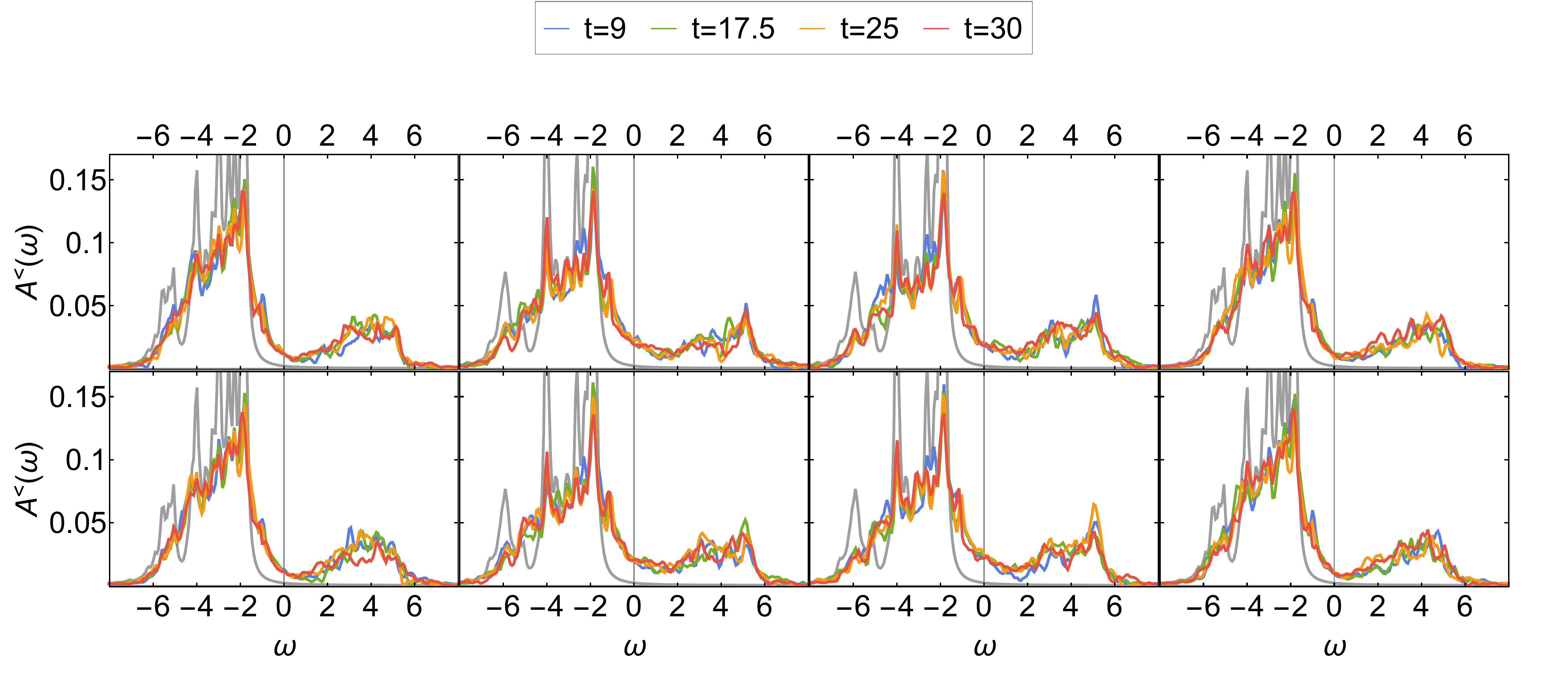}
		 \caption{Site-resolved spectral function $A^<(\omega)$ for $2\times4$ box for the same parameters as in Fig.~\ref{fig:nnn2} and $v_d=0$. The arrangement of the plots corresponds to the arrangement of sites. Different colors correspond to different times after the pulse as shown in the legend at the top. Grey line corresponds to  $A^<(\omega)$ for $t=0$.
			}
		 \label{fig:b24_spec_onsite}
\end{figure*}


In Figs.~\ref{fig:b43_d_sites}-\ref{fig:b26_d_sites} we show site-resolved double occupation for the $4\times3$, $2\times 6$ and $2\times4$ boxes respectively, for the same parameters as the average double occupations shown in Figs.~\ref{fig:dspec1} and~\ref{fig:nnn2}. A similar discussion for the chain geometry can be found in Ref.~\onlinecite{innerberger2020}. For the $4\times 3$ box, Fig.~\ref{fig:b43_d_sites},  the time-dependent double occupation $d(t)$ differs quite strongly between middle (6,7), edge (2,3,5,8,10,11) and corner (1,4,9,12) sites, but all of them show secondary increase of double occupation (impact ionization). For the $2\times6$ box, left panel of Fig.~\ref{fig:b26_d_sites}, the differences between edge (1,6,7,12) and middle (all other sites) are not so large and no sites show impact ionization. In the case of the $2\times 4$ box, right panel of Fig.~\ref{fig:b26_d_sites}, also all sites show impact ionization but it is hardly visible, since both fluctuations of the double occupations and differences between sites are very large. 


For the two systems with stronger boundary effects, the $4\times 3$ and $2\times 4$ boxes, we also show the (lesser) spectral functions $A^<(\omega)$ for all sites independently, for different times after the pulse (Figs.~\ref{fig:b43_spec_onsite} and~\ref{fig:b24_spec_onsite} respectively). The evolution of the spectral weight is analogous to the averaged one in Figs.~\ref{fig:spec1} and~\ref{fig:spec2}, with the spectral weight shift within the upper Hubbard band as described in Sec.~\ref{sec:nn_only}. In both  $4\times 3$ and $2\times 4$ boxes, the edge sites are more insulating than the middle sites, where the gap filling after the pulse is more significant.

\section{Convergence in number of disorder realizations}

\begin{figure}[b]
	\centering
		 \includegraphics[width=\linewidth]{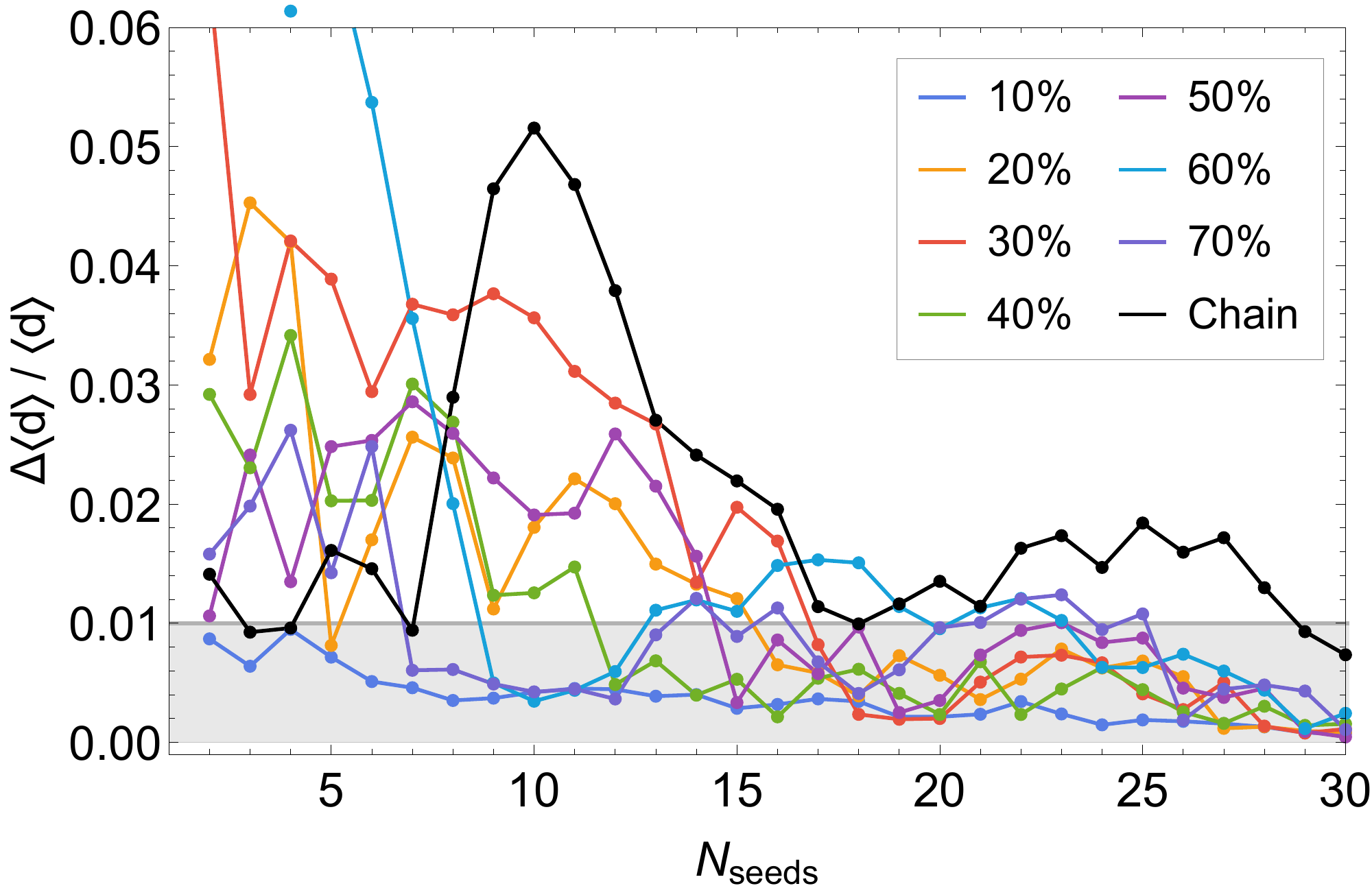}
		 \caption{Difference of $\langle d\rangle(N_{seeds})$  relative to $N_{seeds}=31$ as defined by Eq.~(\ref{eq:erd}) for the {$2 \times 4$} box with different levels of disorder and the $8$-site chain with 30\% disorder. The parameters are: $U=6; \omega=9; a=0.8$. 
			}
		 \label{fig:erd}
\end{figure}

\begin{figure}
		\centering
		 \includegraphics[width=\linewidth]{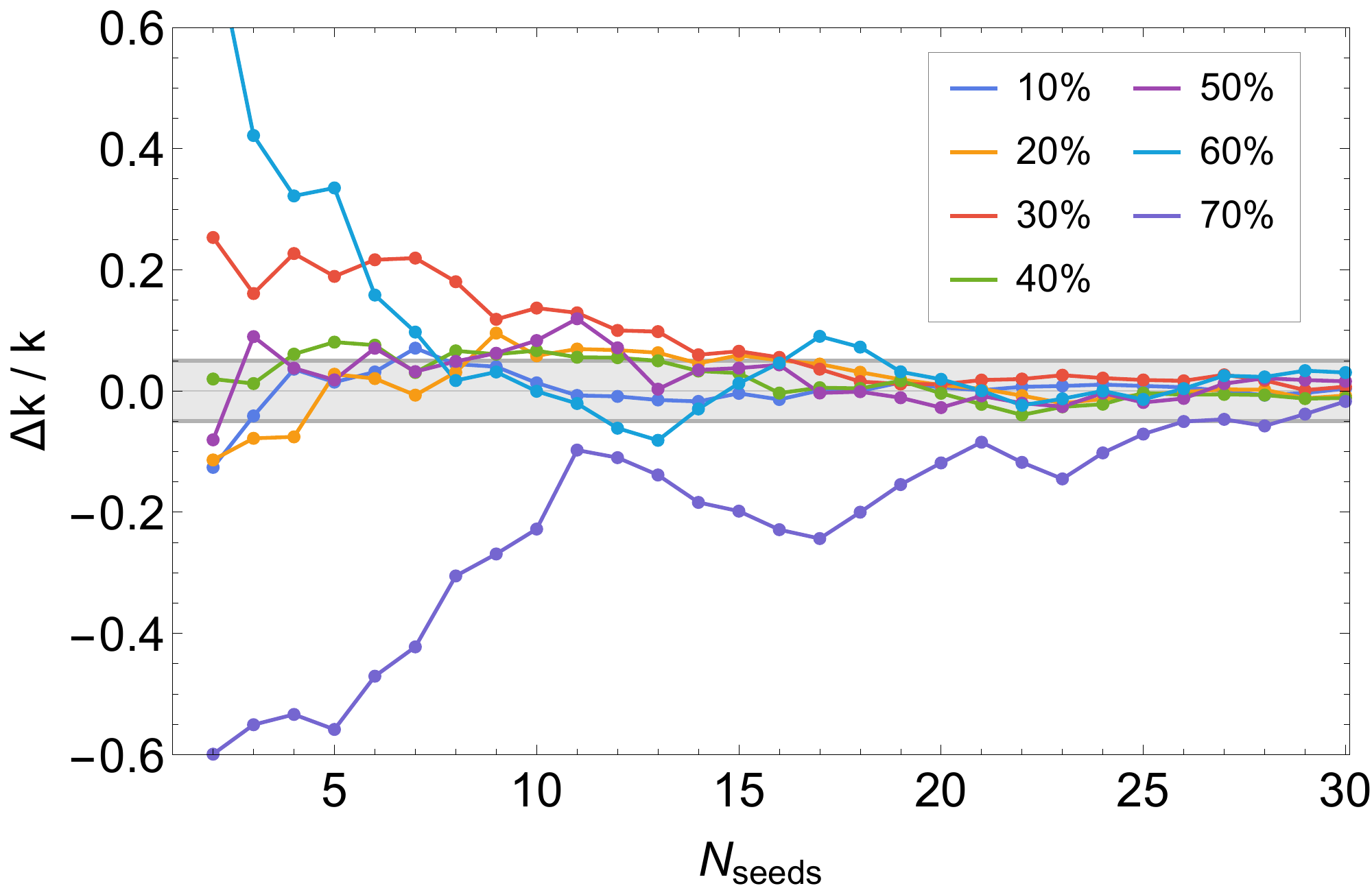}
		 \caption{Difference of  $k(N_{seeds})$  relative to $N_{seeds}=31$ as defined by Eq.~(\ref{eq:erk}) for the {$2 \times 4$} box with different levels of disorder. The same parameters as in Fig.~\ref{fig:erd}.}
		 \label{fig:erk}
\end{figure}
\begin{figure}
				 \includegraphics[width=\linewidth]{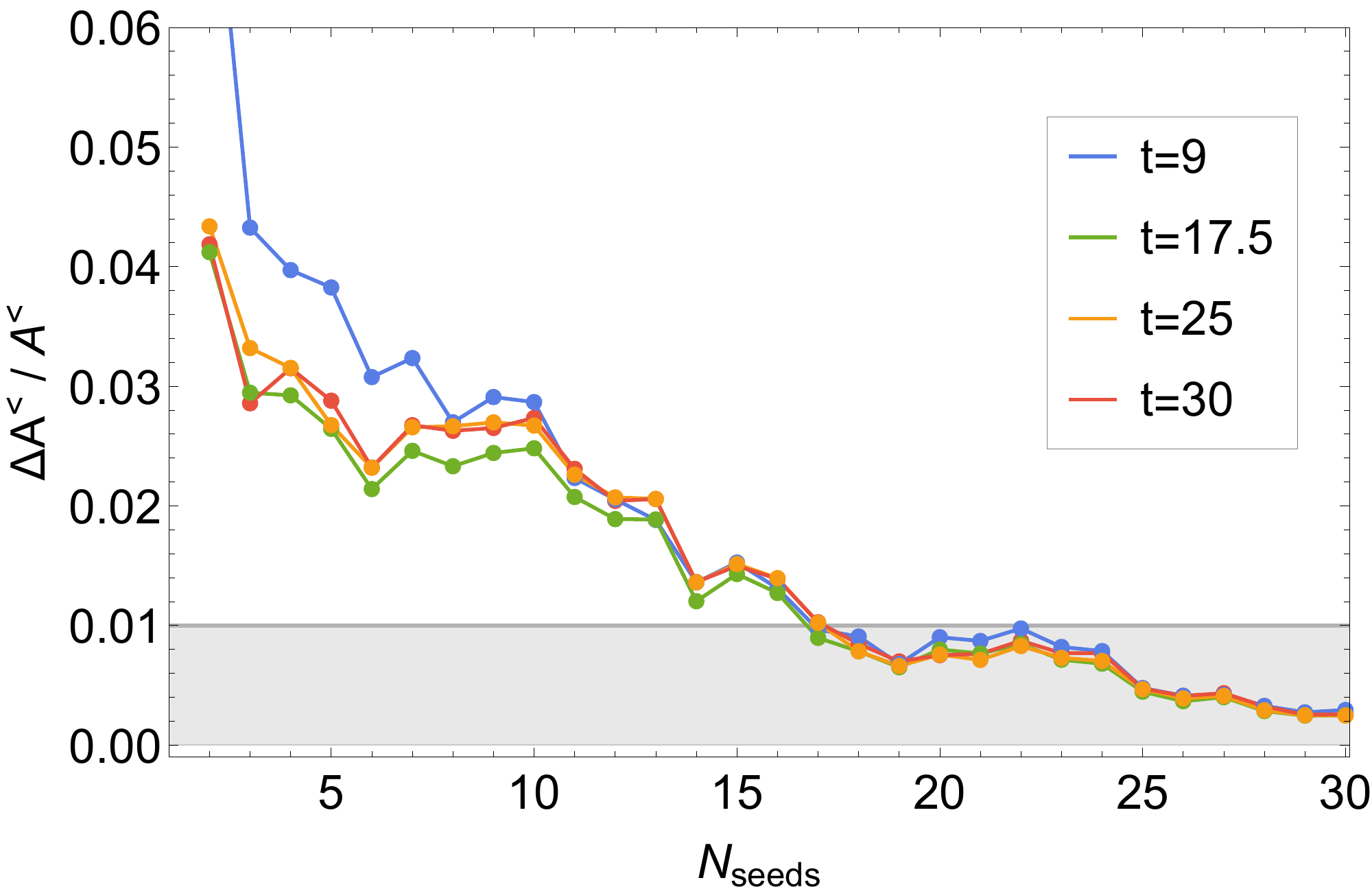}
		 \caption{Difference of  $A_t^<(N_{seeds})$  relative to $N_{seeds}=31$ as defined by Eq.~(\ref{eq:era}) for the {$2 \times 4$} box with 30\% disorder at different times $t$.  The same parameters as in Fig.~\ref{fig:erd}.}
		 \label{fig:era}
\end{figure}

To obtain data presented in Section \ref{Sec:disorder}, we performed disorder averaging over $N_{seeds}=31$ disorder realizations. We here show the convergence of the result for different strength of disorder. Although the number of disorder realisations $N_{seeds}=31$ is relatively small,  our results are reasonably converged. Please keep in mind that besides the disorder averaging, there is also site averaging. This explains why convergence is already achieved with quite few different disorder realizations.

To this end, let us define the relative error for the three quantities studied in the main text. It is given by the relative difference after $N_{seeds}$ disorder realizations to the  $N_{seeds}^{max}=31$ of the main text: 

\noindent (i) relative error in the mean double occupancy (shown in Fig.~\ref{fig:erd}):
	\begin{align}\label{eq:erd}
	&	\frac{\Delta \langle d\rangle}{\langle d\rangle}(N_{seeds})= \nonumber\\
	&	\frac{\int_{0}^{t_{max}}\mid \langle d\rangle(t,N_{seeds}) -\langle d\rangle(t,N_{seeds}^{max})  \mid dt}{\int_{0}^{t_{max}}\langle d\rangle(t,N_{seeds}^{max}) dt}
	\end{align}
\noindent (ii) relative error in the rate of impact ionization (shown in Fig.~\ref{fig:erk}):
	\begin{equation}\label{eq:erk}
		\frac{\Delta k}{k}(N_{seeds})=\frac{k(N_{seeds}) - k(N_{seeds}^{max}) }{k(N_{seeds}^{max})}
	\end{equation}
\noindent (iii) relative error in the average local spectral function (shown in Fig.~\ref{fig:era}):
	\begin{align}\label{eq:era}
	&	\frac{\Delta A_t^<}{ A_t^<}(N_{seeds})= \nonumber \\
	&\frac{\int_{\omega_{min}}^{\omega_{max}}\mid  A_t^<(\omega,N_{seeds}) - A_t^<(\omega,N_{seeds}^{max})  \mid d\omega}{\int_{\omega_{min}}^{\omega_{max}} A_t^<(\omega,N_{seeds}^{max}) d\omega}.
	\end{align}

From Figs. \ref{fig:erd}, \ref{fig:erk} and \ref{fig:era} we see that the for the {$2 \times 4$} box the convergence is rather fast: the relative errors  $\frac{\Delta \langle d\rangle}{\langle d\rangle}(N_{seeds})$ and $\frac{\Delta A_t^<}{ A_t^<}(N_{seeds})$  are below 1\% at all four times  and $\frac{\Delta k}{k}(N_{seeds})$ falls under 5\% error after $N_{seeds} \sim 17$ disorder configurations.

 For the $8$-site chain with 30\% disorder  $\frac{\Delta \langle d\rangle}{\langle d\rangle}(N_{seeds})$ converges somewhat more slowly in Fig. \ref{fig:erd}. We do not calculate $\frac{\Delta k}{k}(N_{seeds})$ for the chain, because for all $N_{seeds}$, $k$ is effectively zero ( $<10^{-6}$).


\section{Eigenvalue degeneracy for systems with disorder or NNN hopping}
\label{app:degeneracies}

\begin{figure}
		\includegraphics[width=\linewidth]{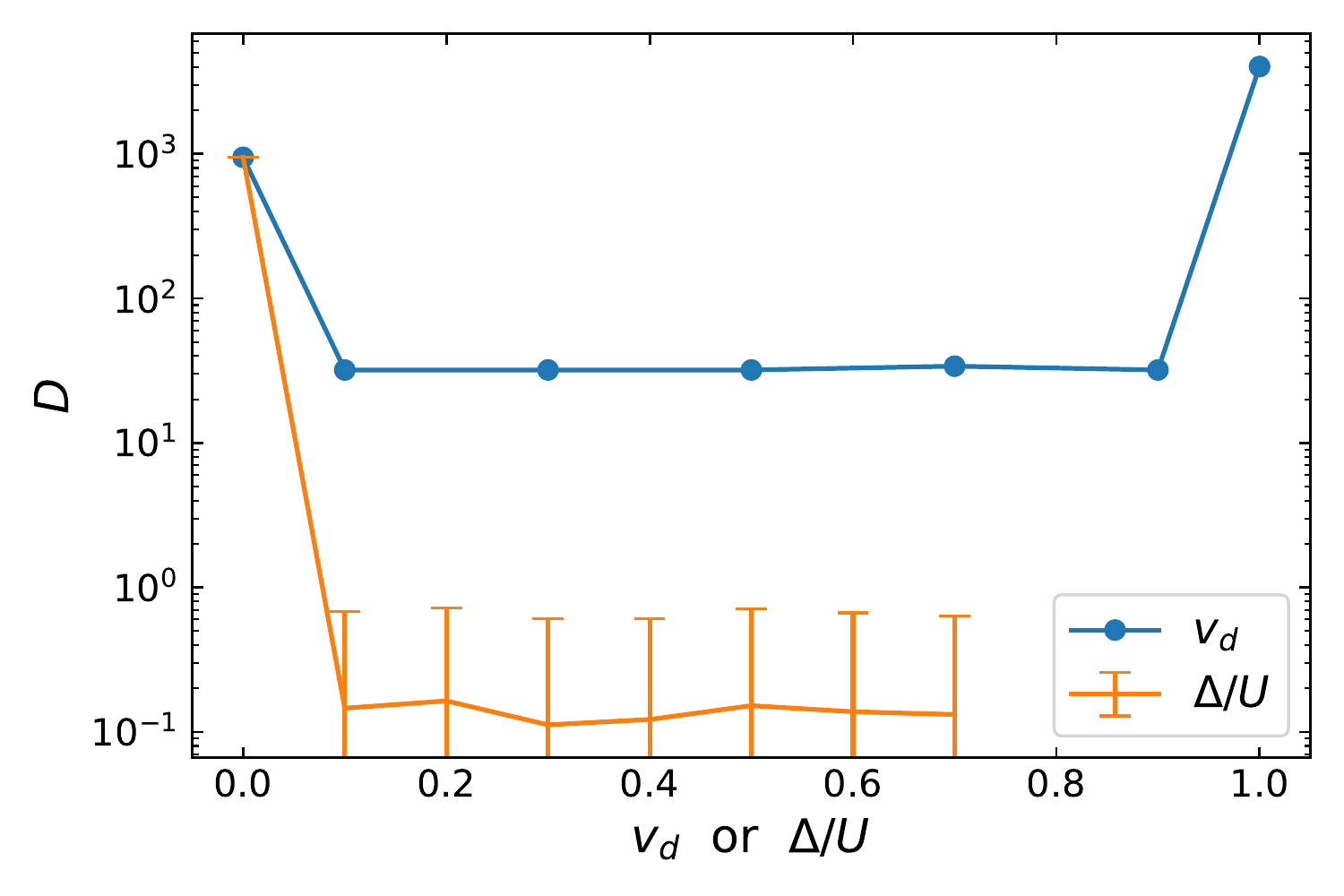}
		\includegraphics[width=\linewidth]{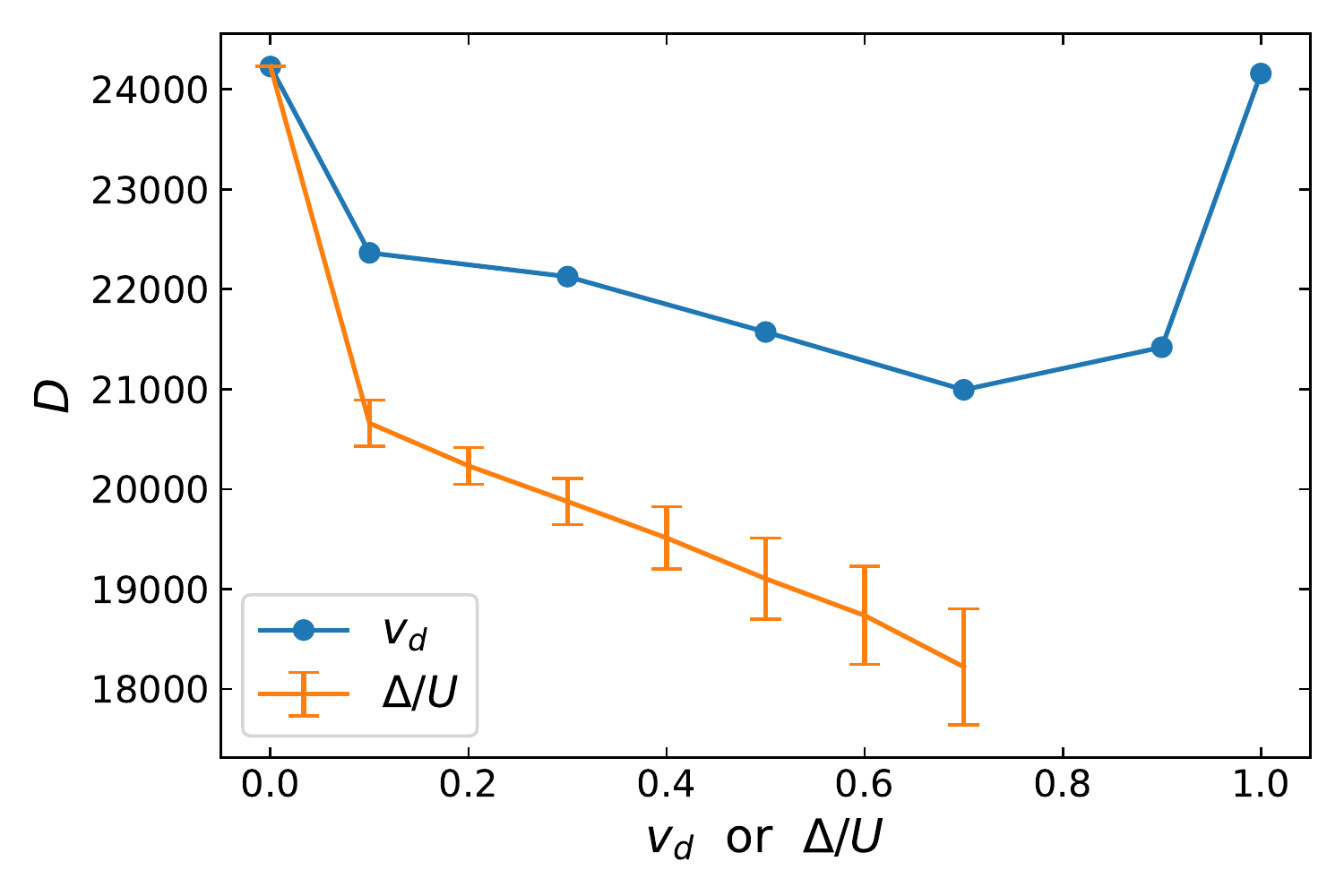}
		 \caption{Measure of degeneracy as defined in Eq.~\eqref{eqn:deg_measure} for a {$2 \times 4$} box depending on  next-neared neighbor hopping $v_d$ (blue) or disorder strength $\Delta/U$ (orange). The numerical tolerance is set to $\delta E=10^{-7}$ for the upper panel (absolute degeneracy within numerical precision) and to $\delta E=0.01$ in the lower panel. The disorder curve is calculated as the average over disorder realizations and the error bars denote standard deviation. Note that   $D={\cal O}(10^{-1})$ in the upper panel for $\Delta/U>0$ is because most disorder realizations have $D=0$ and some (accidental) $D=1$. }
		 \label{fig:degeneracy}
\end{figure}

As a measure of the degree of eigenvalue  degeneracy we consider
\begin{equation}
    D = \sum_{i=1}^{\mathrm{dim}(H)} \sum_{\substack{j=1 \\ j\neq i}}^{\mathrm{dim}(H)} \theta(\delta E -\left| E_i -E_j\right|  ),
    \label{eqn:deg_measure}
\end{equation}
where $E_i$ is the $i$th eigenenergy of the  Hamiltonian \eqref{eq:Hubbard} (in the block with a fixed particle number and spin). $\delta E$ is a tolerance width regarding differences in eigenvalues. For  $ \delta E \rightarrow 0$ (or machine precision), $D$ measures the actual degeneracies of the system. We chose this measure since it takes the multiplicity of the degeneracy into account.

In the $2\times 4$ box any amount of disorder lifts all degeneracies in the system (see top panel of Fig.~\ref{fig:degeneracy}) and therefore we rather measure how many eigenvalues are clustered within a small interval $\delta E =0.01$ (bottom panel), which counts not exact but near-degeneracies. As we see in the bottom panel of Fig.~\ref{fig:degeneracy}, $D$ is reduced with increasing disorder strength $\Delta/U$ and NNN hopping.  In the case of NNN hopping, the number of degeneracies increases when we set $v_d=1$, which makes the NN and NNN hoppings equal and the system more symmetric again.

\section{Parameter scan for the $2\times4$ box}
\label{app:scan}

\begin{figure}[b]
	\centering
		 \includegraphics[width=0.8\linewidth]{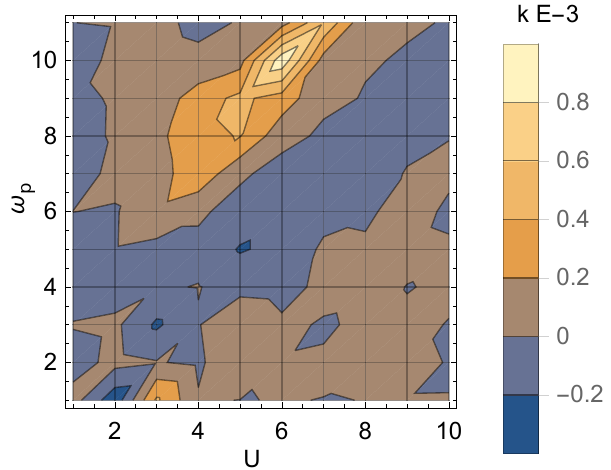}
		 \caption{Dependence of impact ionization rate $k$ (as defined in Sec.~\ref{Sec:NNN}) on the parameters $U$ and pulse frequency $\omega_p$ for the $2\times4$ box with only NN hopping ($v_d=0$).
			}
		 \label{fig:kd_vs_U_omega}
\end{figure}

In Fig.~\ref{fig:kd_vs_U_omega} we show the dependence of the impact ionization rate $k$ (as defined in Sec.~\ref{Sec:NNN}) on the parameters $U$ and pulse frequency $\omega_p$ for the $2\times4$ box with only NN hopping ($v_d=0$; that is the reference system to which we add NNN hopping and disorder). We see that the range of parameters, where we see impact ionization (positive $k$) is quite small, with a maximum at $U=6$ and $\omega_p\approx 10$. For the computations in the paper we chose a slightly smaller $\omega_p=9$ to avoid saturation of the double occupation already in the reference system. The same parameters we choose then for all $2\times4$ systems. In the case of $12$-site systems we chose the parameters based on Ref.~\onlinecite{note_parameters}.



\end{document}